\documentclass[preprint2]{aastex}

\begin{document}

\title{ADDENDUM: AN INVESTIGATION INTO THE PROMINENCE OF SPIRAL GALAXY BULGES [AJ, 121, 820 (2001)]}
\author{Alister W. Graham}
\affil{Department of Astronomy, University of Florida, P.O.\ Box 112055, Gainesville, FL 32611, USA}
\email{Graham@astro.ufl.edu}

\begin{abstract}
The best-fitting $B, R, I,$ and $K$-band (S\'ersic $R^{1/n}$ bulge +
exponential disk) model parameters for the de Jong \& van der Kruit (1994)
sample of 86 spiral galaxies are made electronically available.  
To avoid large biases from likely nuclear star clusters, 
the central data point has now been excluded from a number of galaxies. 
This sample covers all spiral galaxy types and provides the
largest homogeneous set of structural parameters for
spiral galaxies modeled with an $R^{1/n}$ bulge.  Readers are welcome 
to request diagrams of the fitted profiles from the Author. 
\end{abstract}

\section{ }

\begin{deluxetable}{crrrrrrccc}
\tablecolumns{10}
\tablewidth{0pc}
\tablecaption{$K$-band model parameters}
\tablehead{
\colhead{Galaxy} & \colhead{$\mu_{0,d}$} & \colhead{$h$}   &  \colhead{$\mu_{e,b}$}  
& \colhead{$R_{e,b}$} & \colhead{$n$}   & 
\colhead{extinc} & \colhead{Type} & \colhead{Dist.} & \colhead{$b/a$} \\
\colhead{UGC \#} & \colhead{mag/$\sq\arcsec$} & \colhead{arcsec}   &  
\colhead{mag/$\sq\arcsec$}  & \colhead{arcsec} & \colhead{}   & 
\colhead{$K$-mag} & \colhead{} & \colhead{km s$^{-1}$} & \colhead{}
}
\startdata
00089 & \multicolumn{5}{c}{no fit}                                                                                                                               &   0.014 &  1 &  4566 &  0.67 \\
00093 &  18.55$^{+  0.04}_{-  0.19}$ &  15.79$^{+  1.36}_{-  2.87}$ &  18.91$^{+  0.09}_{-  0.31}$ &   1.89$^{+  0.14}_{-  0.43}$ &   0.64$^{+  0.09}_{-  0.35}$ &   0.021 &  8 &  4951 &  0.64 \\
00242 &  17.44$^{+  0.01}_{-  0.00}$ &  12.71$^{+  0.26}_{-  0.26}$ &  17.76$^{+  0.01}_{-  0.01}$ &   1.53$^{+  0.01}_{-  0.02}$ &   0.21$^{+  0.02}_{-  0.01}$ &   0.024 &  7 &  4395 &  0.77 \\
00334 &  19.97$^{+  0.03}_{-  0.07}$ &  13.52$^{+  1.93}_{-  2.01}$ &  20.73$^{+  0.07}_{-  0.04}$ &   2.82$^{+  0.08}_{-  0.11}$ &   0.28$^{+  0.03}_{-  0.05}$ &   0.020 &  9 &  4627 &  0.76 \\
00438 &  16.81$^{+  0.25}_{-  0.43}$ &  13.11$^{+  0.00}_{-  1.35}$ &  17.87$^{+  0.85}_{-  0.81}$ &   8.80$^{+  7.40}_{-  4.34}$ &   2.70$^{+  0.67}_{-  0.50}$ &   0.013 &  5 &  4534 &  0.78 \\
00463 &  16.65$^{+  0.10}_{-  0.20}$ &  10.94$^{+  0.77}_{-  1.13}$ &  16.40$^{+  0.06}_{-  0.07}$ &   1.86$^{+  0.12}_{-  0.20}$ &   0.40$^{+  0.08}_{-  0.12}$ &   0.033 &  5 &  4452 &  0.83 \\
00490 &  16.77$^{+  0.02}_{-  0.06}$ &  12.19$^{+  0.30}_{-  0.49}$ &  17.79$^{+  0.07}_{-  0.28}$ &   3.12$^{+  0.21}_{-  0.53}$ &   1.25$^{+  0.06}_{-  0.26}$ &   0.018 &  5 &  4552 &  0.71 \\
00508 &  17.35$^{+  0.07}_{-  0.08}$ &  22.39$^{+  1.20}_{-  1.18}$ &  15.82$^{+  0.05}_{-  0.04}$ &   4.23$^{+  0.15}_{-  0.13}$ &   1.36$^{+  0.06}_{-  0.02}$ &   0.025 &  2 &  4661 &  0.89 \\
00628 &  19.77$^{+  0.02}_{-  0.04}$ &  12.01$^{+  1.44}_{-  1.45}$ &  21.05$^{+  0.07}_{-  0.04}$ &   2.72$^{+  0.10}_{-  0.12}$ &   0.53$^{+  0.01}_{-  0.02}$ &   0.016 &  9 &  5446 &  0.63 \\
01305 &  17.84$^{+  0.08}_{-  0.07}$ &  29.01$^{+  1.52}_{-  1.48}$ &  17.89$^{+  0.11}_{-  0.11}$ &   7.97$^{+  0.72}_{-  0.66}$ &   1.83$^{+  0.10}_{-  0.10}$ &   0.026 &  4 &  2665 &  0.81 \\
01455 &  17.57$^{+  0.19}_{-  0.21}$ &  15.80$^{+  1.71}_{-  1.74}$ &  16.25$^{+  0.30}_{-  0.17}$ &   2.91$^{+  0.48}_{-  0.32}$ &   1.48$^{+  0.34}_{-  0.15}$ &   0.038 &  4 &  5117 &  0.94 \\
01551 &  18.95$^{+  0.05}_{-  0.10}$ &  23.69$^{+  2.81}_{-  3.24}$ &  19.84$^{+  0.01}_{-  0.00}$ &   2.85$^{+  0.29}_{-  0.43}$ &   0.68$^{+  0.03}_{-  0.02}$ &   0.033 &  8 &  2671 &  0.89 \\
01559 &  20.14$^{+  0.18}_{-  0.14}$ &  20.45$^{+ 11.60}_{-  6.41}$ &  19.97$^{+  0.04}_{-  0.00}$ &   5.83$^{+  0.54}_{-  0.14}$ &   0.72$^{+  0.07}_{-  0.01}$ &   0.023 &  7 &  3619 &  0.87 \\
01577 &  17.80$^{+  0.25}_{-  0.26}$ &  15.24$^{+  2.31}_{-  2.10}$ &  16.87$^{+  0.48}_{-  0.35}$ &   3.17$^{+  1.04}_{-  0.56}$ &   1.78$^{+  0.48}_{-  0.52}$ &   0.021 &  4 &  5276 &  0.79 \\
01719 &  18.14$^{+  0.56}_{-  0.65}$ &  16.14$^{+  4.00}_{-  3.66}$ &  18.40$^{+  0.53}_{-  1.14}$ &   7.08$^{+  4.05}_{-  3.89}$ &   2.32$^{+  0.29}_{-  0.99}$ &   0.030 &  3 &  8213 &  0.74 \\
01792 &  17.46$^{+  0.08}_{-  0.10}$ &  14.40$^{+  1.15}_{-  1.27}$ &  17.14$^{+  0.09}_{-  0.06}$ &   2.35$^{+  0.15}_{-  0.15}$ &   0.61$^{+  0.10}_{-  0.06}$ &   0.033 &  5 &  4987 &  0.59 \\
02064 &  18.09$^{+  0.06}_{-  0.08}$ &  17.75$^{+  1.19}_{-  1.29}$ &  17.02$^{+  0.09}_{-  0.12}$ &   1.96$^{+  0.12}_{-  0.14}$ &   1.38$^{+  0.07}_{-  0.09}$ &   0.048 &  4 &  4265 &  0.70 \\
02081 &  19.48$^{+  0.26}_{-  0.13}$ &  19.48$^{+  6.81}_{-  3.72}$ &  20.22$^{+  0.64}_{-  0.11}$ &   4.27$^{+  3.32}_{-  0.72}$ &   1.15$^{+  0.52}_{-  0.09}$ &   0.010 &  6 &  2616 &  0.60 \\
02124 &  17.92$^{+  0.09}_{-  0.13}$ &  19.50$^{+  2.14}_{-  2.24}$ &  16.42$^{+  0.01}_{-  0.02}$ &   4.46$^{+  0.06}_{-  0.09}$ &   0.75$^{+  0.04}_{-  0.05}$ &   0.012 &  0 &  2631 &  0.88 \\
02125 & \multicolumn{5}{c}{no data}                                                                                                                              &   0.052 &  5 &  5187 &  0.96 \\
02197 &  17.91$^{+  0.28}_{-  0.19}$ &  11.95$^{+  2.76}_{-  1.77}$ &  18.42$^{+  0.62}_{-  0.06}$ &   2.52$^{+  1.67}_{-  0.37}$ &   0.52$^{+  0.56}_{-  0.16}$ &   0.079 &  6 &  5098 &  0.62 \\
02368 &  17.74$^{+  0.17}_{-  0.18}$ &  14.91$^{+  1.73}_{-  1.70}$ &  15.03$^{+  0.10}_{-  0.21}$ &   1.46$^{+  0.06}_{-  0.13}$ &   3.20$^{+  2.19}_{-  1.10}$ &   0.061 &  3 &  3568 &  0.92 \\
02595 & \multicolumn{5}{c}{no data}                                                                                                                              &   0.063 &  4 &  5904 &  0.78 \\
03066 &  17.11$^{+  0.16}_{-  0.30}$ &  11.31$^{+  1.46}_{-  1.93}$ &  17.33$^{+  0.11}_{-  0.19}$ &   2.10$^{+  0.33}_{-  0.51}$ &   0.60$^{+  0.12}_{-  0.26}$ &   0.113 &  7 &  4640 &  0.73 \\
03080 &  18.16$^{+  0.10}_{-  0.03}$ &  14.42$^{+  1.84}_{-  0.90}$ &  18.40$^{+  0.09}_{-  0.03}$ &   1.60$^{+  0.20}_{-  0.06}$ &   0.38$^{+  0.12}_{-  0.04}$ &   0.032 &  5 &  3541 &  0.91 \\
03140 &  17.40$^{+  0.08}_{-  0.08}$ &  11.25$^{+  0.04}_{-  0.13}$ &  18.71$^{+  0.34}_{-  0.32}$ &   9.46$^{+  2.36}_{-  1.78}$ &   3.88$^{+  0.32}_{-  0.29}$ &   0.030 &  5 &  4633 &  0.94 \\
04126 &  17.84$^{+  0.29}_{-  0.17}$ &  17.72$^{+  2.06}_{-  1.65}$ &  16.95$^{+  1.45}_{-  0.33}$ &   2.95$^{+  3.40}_{-  0.46}$ &   2.06$^{+  2.28}_{-  0.58}$ &   0.019 &  3 &  4841 &  0.85 \\
04256 &  17.43$^{+  0.12}_{-  0.11}$ &  16.51$^{+  1.13}_{-  1.07}$ &  17.63$^{+  0.42}_{-  0.37}$ &   4.53$^{+  1.39}_{-  0.94}$ &   2.55$^{+  0.39}_{-  0.36}$ &   0.020 &  5 &  5260 &  0.86 \\
04308 &  17.64$^{+  0.05}_{-  0.06}$ &  16.30$^{+  0.81}_{-  0.87}$ &  16.99$^{+  0.06}_{-  0.06}$ &   2.06$^{+  0.09}_{-  0.09}$ &   0.75$^{+  0.06}_{-  0.06}$ &   0.014 &  5 &  3566 &  0.77 \\
04368 &  17.84$^{+  0.05}_{-  0.07}$ &  14.83$^{+  0.89}_{-  0.97}$ &  18.62$^{+  0.00}_{-  0.00}$ &   2.49$^{+  0.14}_{-  0.18}$ &   0.48$^{+  0.03}_{-  0.04}$ &   0.014 &  6 &  3870 &  0.80 \\
04375 &  17.20$^{+  0.12}_{-  0.03}$ &  16.70$^{+  1.34}_{-  0.47}$ &  18.65$^{+  0.44}_{-  0.17}$ &   3.26$^{+  2.10}_{-  0.30}$ &   1.12$^{+  0.32}_{-  0.23}$ &   0.016 &  5 &  2061 &  0.66 \\
04422 &  18.40$^{+  0.03}_{-  0.02}$ &  27.05$^{+  1.12}_{-  1.13}$ &  16.47$^{+  0.01}_{-  0.02}$ &   3.57$^{+  0.02}_{-  0.03}$ &   0.90$^{+  0.01}_{-  0.02}$ &   0.015 &  4 &  4330 &  0.82 \\
04458 &  18.57$^{+  0.18}_{-  0.21}$ &  22.75$^{+  2.18}_{-  2.18}$ &  16.40$^{+  0.10}_{-  0.12}$ &   5.30$^{+  0.30}_{-  0.36}$ &   2.09$^{+  0.11}_{-  0.14}$ &   0.013 &  1 &  4741 &  0.87 \\
05103 &  16.49$^{+  0.07}_{-  0.08}$ &  13.94$^{+  0.62}_{-  0.72}$ &  15.95$^{+  0.07}_{-  0.06}$ &   2.23$^{+  0.10}_{-  0.13}$ &   0.69$^{+  0.05}_{-  0.07}$ &   0.010 &  3 &  3730 &  0.61 \\
05303 &  17.86$^{+  0.03}_{-  0.02}$ &  36.75$^{+  1.57}_{-  1.59}$ &  17.57$^{+  0.02}_{-  0.01}$ &   5.34$^{+  0.09}_{-  0.10}$ &   0.78$^{+  0.01}_{-  0.02}$ &   0.013 &  5 &  1408 &  0.59 \\
05510 &  17.51$^{+  0.20}_{-  0.14}$ &  19.41$^{+  1.54}_{-  1.30}$ &  17.57$^{+  1.17}_{-  0.59}$ &   3.94$^{+  4.06}_{-  1.12}$ &   2.06$^{+  1.13}_{-  0.62}$ &   0.009 &  4 &  1298 &  0.81 \\
05554 &  17.03$^{+  0.04}_{-  0.05}$ &  16.80$^{+  0.53}_{-  0.56}$ &  15.79$^{+  0.03}_{-  0.03}$ &   2.70$^{+  0.05}_{-  0.06}$ &   0.94$^{+  0.03}_{-  0.03}$ &   0.010 &  1 &  1217 &  0.69 \\
05633 &  19.90$^{+  0.14}_{-  0.20}$ &  27.13$^{+  6.04}_{-  5.43}$ &  20.86$^{+  0.21}_{-  0.40}$ &   6.78$^{+  2.06}_{-  2.23}$ &   1.14$^{+  0.17}_{-  0.33}$ &   0.016 &  8 &  1383 &  0.63 \\
05842 &  18.30$^{+  0.00}_{-  0.00}$ &  36.88$^{+  0.32}_{-  0.33}$ &  18.49$^{+  0.00}_{-  0.00}$ &   3.33$^{+  0.01}_{-  0.00}$ &   0.69$^{+  0.01}_{-  0.00}$ &   0.010 &  6 &  1260 &  0.85 \\
06028 &  \multicolumn{5}{c}{no fit}                                                                                                                              &   0.012 &  3 &  1102 &  0.58 \\
06077 &  17.51$^{+  0.01}_{-  0.01}$ &  17.88$^{+  0.22}_{-  0.23}$ &  17.22$^{+  0.01}_{-  0.01}$ &   2.39$^{+  0.02}_{-  0.03}$ &   0.71$^{+  0.00}_{-  0.01}$ &   0.008 &  3 &  1434 &  0.82 \\
06123 &  17.22$^{+  0.12}_{-  0.20}$ &  23.26$^{+  2.09}_{-  2.83}$ &  16.10$^{+  0.22}_{-  0.30}$ &   2.61$^{+  0.37}_{-  0.47}$ &   1.78$^{+  0.22}_{-  0.24}$ &   0.009 &  3 &  0979 &  0.93 \\
06277 &  17.16$^{+  0.09}_{-  0.11}$ &  20.02$^{+  1.51}_{-  1.61}$ &  16.86$^{+  0.13}_{-  0.09}$ &   3.43$^{+  0.39}_{-  0.30}$ &   1.20$^{+  0.10}_{-  0.08}$ &   0.008 &  5 &  1193 &  0.88 \\
06445 &  16.71$^{+  0.06}_{-  0.05}$ &  13.14$^{+  0.51}_{-  0.52}$ &  16.11$^{+  0.03}_{-  0.02}$ &   3.42$^{+  0.08}_{-  0.09}$ &   0.87$^{+  0.03}_{-  0.02}$ &   0.010 &  4 &  1239 &  0.84 \\
06453 &  16.96$^{+  0.06}_{-  0.05}$ &  14.70$^{+  0.37}_{-  0.33}$ &  18.30$^{+  0.08}_{-  0.08}$ &   6.91$^{+  0.75}_{-  0.62}$ &   1.45$^{+  0.06}_{-  0.06}$ &   0.009 &  4 &  1163 &  0.75 \\
06460 &  17.05$^{+  0.06}_{-  0.07}$ &  22.26$^{+  1.11}_{-  1.29}$ &  16.61$^{+  0.09}_{-  0.13}$ &   2.16$^{+  0.17}_{-  0.19}$ &   1.18$^{+  0.07}_{-  0.09}$ &   0.009 &  4 &  1156 &  0.75 \\
06536 &  19.33$^{+  0.16}_{-  0.23}$ &  26.85$^{+  3.83}_{-  3.88}$ &  16.75$^{+  0.07}_{-  0.09}$ &   5.23$^{+  0.19}_{-  0.28}$ &   1.89$^{+  0.08}_{-  0.11}$ &   0.009 &  3 &  6970 &  0.62 \\
06693 &  18.01$^{+  0.02}_{-  0.03}$ &  16.80$^{+  0.61}_{-  0.63}$ &  17.06$^{+  0.02}_{-  0.04}$ &   1.45$^{+  0.03}_{-  0.03}$ &   0.70$^{+  0.03}_{-  0.03}$ &   0.009 &  4 &  6909 &  0.85 \\
06746 &  17.60$^{+  0.26}_{-  0.27}$ &  17.86$^{+  2.06}_{-  2.04}$ &  16.44$^{+  0.69}_{-  0.52}$ &   3.41$^{+  1.57}_{-  0.75}$ &   2.33$^{+  0.81}_{-  0.93}$ &   0.009 &  0 &  6946 &  0.76 \\
06754 &  18.74$^{+  0.12}_{-  0.15}$ &  25.55$^{+  2.89}_{-  2.89}$ &  17.56$^{+  0.28}_{-  0.34}$ &   4.35$^{+  0.73}_{-  0.73}$ &   3.11$^{+  0.33}_{-  0.42}$ &   0.010 &  3 &  7025 &  0.89 \\
07169 &  16.17$^{+  0.03}_{-  0.03}$ &   8.39$^{+  0.17}_{-  0.16}$ &  16.29$^{+  0.06}_{-  0.05}$ &   1.91$^{+  0.08}_{-  0.08}$ &   0.96$^{+  0.06}_{-  0.06}$ &   0.012 &  5 &  2167 &  0.81 \\
07315 &  16.08$^{+  0.04}_{-  0.04}$ &  14.77$^{+  0.32}_{-  0.32}$ &  16.90$^{+  0.10}_{-  0.06}$ &   2.84$^{+  0.24}_{-  0.19}$ &   0.72$^{+  0.09}_{-  0.06}$ &   0.011 &  4 &  0867 &  0.67 \\
07450 &  17.61$^{+  0.22}_{-  0.13}$ &  55.00$^{+  6.90}_{-  4.89}$ &  16.27$^{+  0.59}_{-  0.25}$ &   8.83$^{+  3.24}_{-  1.03}$ &   1.62$^{+  0.61}_{-  0.26}$ &   0.010 &  4 &  1571 &  0.89 \\
07523 &  17.50$^{+  0.12}_{-  0.16}$ &  30.11$^{+  2.67}_{-  2.75}$ &  16.05$^{+  0.13}_{-  0.16}$ &   5.68$^{+  0.48}_{-  0.50}$ &   1.80$^{+  0.13}_{-  0.16}$ &   0.011 &  3 &  0922 &  0.81 \\
07594 &  17.06$^{+  0.14}_{-  0.15}$ &  46.34$^{+  3.76}_{-  3.60}$ &  16.90$^{+  0.24}_{-  0.26}$ &  13.92$^{+  2.53}_{-  2.20}$ &   2.41$^{+  0.20}_{-  0.22}$ &   0.010 &  2 &  1954 &  0.69 \\
07876 &  18.33$^{+  0.07}_{-  0.11}$ &  19.18$^{+  1.38}_{-  1.72}$ &  19.58$^{+  0.02}_{-  0.03}$ &   2.59$^{+  0.39}_{-  0.54}$ &   0.56$^{+  0.06}_{-  0.10}$ &   0.010 &  7 &  0960 &  0.73 \\
07901 &  15.65$^{+  0.00}_{-  0.01}$ &  15.78$^{+  0.09}_{-  0.10}$ &  15.79$^{+  0.00}_{-  0.01}$ &   3.07$^{+  0.01}_{-  0.02}$ &   0.51$^{+  0.01}_{-  0.00}$ &   0.010 &  5 &  0805 &  0.67 \\
08279 & \multicolumn{5}{c}{no fit}                                                                                                                               &   0.005 &  5 &  2612 &  0.77 \\
08289 &  17.74$^{+  0.08}_{-  0.14}$ &  19.24$^{+  1.12}_{-  1.59}$ &  15.59$^{+  0.04}_{-  0.05}$ &   3.06$^{+  0.05}_{-  0.07}$ &   0.68$^{+  0.04}_{-  0.05}$ &   0.009 &  4 &  3362 &  0.89 \\
08865 &  18.35$^{+  0.21}_{-  0.28}$ &  29.75$^{+  4.69}_{-  4.69}$ &  16.98$^{+  0.19}_{-  0.21}$ &   6.42$^{+  0.78}_{-  0.83}$ &   1.92$^{+  0.17}_{-  0.19}$ &   0.008 &  2 &  2386 &  0.76 \\
09024 &  21.69$^{+  0.00}_{-  0.65}$ &  24.39$^{+ 12.48}_{- 12.94}$ &  19.37$^{+  0.00}_{-  0.18}$ &   4.09$^{+  0.03}_{-  0.43}$ &   1.18$^{+  0.00}_{-  0.20}$ &   0.010 &  ? &  2323 &  0.90 \\
09061 &  18.83$^{+  0.02}_{-  0.04}$ &  42.25$^{+  2.57}_{-  2.64}$ &  16.52$^{+  0.02}_{-  0.02}$ &   3.93$^{+  0.03}_{-  0.04}$ &   1.18$^{+  0.03}_{-  0.03}$ &   0.009 &  4 &  5443 &  0.91 \\
09481 &  17.69$^{+  0.03}_{-  0.15}$ &  15.65$^{+  0.46}_{-  1.36}$ &  17.62$^{+  0.06}_{-  0.22}$ &   2.72$^{+  0.10}_{-  0.41}$ &   0.97$^{+  0.05}_{-  0.20}$ &   0.006 &  4 &  3742 &  0.74 \\
09915 &  17.30$^{+  0.07}_{-  0.08}$ &  15.90$^{+  0.74}_{-  0.82}$ &  16.87$^{+  0.07}_{-  0.08}$ &   2.89$^{+  0.17}_{-  0.19}$ &   1.18$^{+  0.06}_{-  0.05}$ &   0.016 &  3 &  1827 &  0.90 \\
09926 &  17.13$^{+  0.53}_{-  0.60}$ &  18.11$^{+  0.34}_{-  2.00}$ &  17.29$^{+  0.34}_{-  0.39}$ &  12.25$^{+  4.32}_{-  4.04}$ &   2.16$^{+  0.23}_{-  0.21}$ &   0.020 &  5 &  1958 &  0.72 \\
09943 &  17.29$^{+  0.21}_{-  0.20}$ &  20.05$^{+  0.81}_{-  0.84}$ &  18.00$^{+  0.09}_{-  0.09}$ &  15.14$^{+  1.95}_{-  1.89}$ &   1.86$^{+  0.06}_{-  0.06}$ &   0.015 &  5 &  1957 &  0.64 \\
10083 &  17.01$^{+  0.03}_{-  0.03}$ &  15.55$^{+  0.36}_{-  0.36}$ &  18.87$^{+  0.01}_{-  0.02}$ &   5.07$^{+  0.27}_{-  0.29}$ &   0.92$^{+  0.00}_{-  0.01}$ &   0.013 &  2 &  1854 &  0.82 \\
10437 & \multicolumn{5}{c}{NP, no fit}                                                                                                                           &   0.003 &  7 &  2596 &  0.92 \\
10445 &  18.88$^{+  0.11}_{-  0.19}$ &  18.43$^{+  3.78}_{-  3.80}$ &  19.89$^{+  0.12}_{-  0.01}$ &   4.99$^{+  0.56}_{-  0.71}$ &   0.54$^{+  0.08}_{-  0.10}$ &   0.012 &  6 &  0963 &  0.78 \\
10584 &  18.16$^{+  0.03}_{-  0.04}$ &  18.38$^{+  1.11}_{-  1.15}$ &  17.24$^{+  0.02}_{-  0.02}$ &   2.18$^{+  0.03}_{-  0.04}$ &   0.62$^{+  0.02}_{-  0.02}$ &   0.005 &  5 &  5260 &  0.88 \\
11628 &  18.90$^{+  0.17}_{-  0.27}$ &  43.84$^{+  8.98}_{-  8.25}$ &  17.57$^{+  0.15}_{-  0.23}$ &  12.18$^{+  1.16}_{-  1.61}$ &   3.40$^{+  0.15}_{-  0.22}$ &   0.036 &  2 &  4211 &  0.70 \\
11708 &     NP                       &  13.69$^{+  0.59}_{-  0.60}$ &      NP                      &   2.03$^{+  0.08}_{-  0.09}$ &   0.98$^{+  0.06}_{-  0.07}$ &   0.025 &  5 &  4163 &  0.81 \\
11872 &  15.90$^{+  0.28}_{-  0.29}$ &  13.46$^{+  1.17}_{-  1.11}$ &  15.87$^{+  0.15}_{-  0.28}$ &   6.48$^{+  1.09}_{-  1.31}$ &   1.72$^{+  0.08}_{-  0.24}$ &   0.026 &  3 &  1150 &  0.74 \\
12151 &  20.10$^{+  0.05}_{-  0.11}$ &  21.12$^{+  4.80}_{-  4.67}$ &  20.92$^{+  0.03}_{-  0.02}$ &   4.86$^{+  0.46}_{-  0.50}$ &   0.60$^{+  0.08}_{-  0.07}$ &   0.025 & 10 &  1755 &  0.87 \\
12343 &  16.90$^{+  0.08}_{-  0.11}$ &  29.15$^{+  2.05}_{-  2.33}$ &  16.41$^{+  0.06}_{-  0.07}$ &   5.28$^{+  0.31}_{-  0.40}$ &   1.14$^{+  0.04}_{-  0.05}$ &   0.041 &  5 &  2381 &  0.77 \\
12379 &  17.76$^{+  0.30}_{-  0.31}$ &  17.01$^{+  2.61}_{-  2.43}$ &  16.75$^{+  0.61}_{-  0.53}$ &   3.85$^{+  1.60}_{-  0.98}$ &   2.60$^{+  0.68}_{-  0.61}$ &   0.031 &  4 &  6213 &  0.91 \\
12391 &  17.78$^{+  0.11}_{-  0.11}$ &  13.52$^{+  1.79}_{-  1.43}$ &  17.21$^{+  0.20}_{-  0.13}$ &   1.26$^{+  0.15}_{-  0.10}$ &   0.57$^{+  0.20}_{-  0.14}$ &   0.032 &  5 &  4887 &  0.89 \\
12511 &  18.20$^{+  0.04}_{-  0.05}$ &  12.09$^{+  0.58}_{-  0.60}$ &  18.10$^{+  0.05}_{-  0.04}$ &   2.20$^{+  0.08}_{-  0.08}$ &   0.70$^{+  0.05}_{-  0.04}$ &   0.015 &  6 &  3554 &  0.82 \\
12614 &  17.24$^{+  0.02}_{-  0.03}$ &  19.64$^{+  0.64}_{-  0.67}$ &  15.52$^{+  0.03}_{-  0.03}$ &   1.71$^{+  0.03}_{-  0.03}$ &   1.37$^{+  0.03}_{-  0.04}$ &   0.018 &  5 &  3489 &  0.74 \\
12638 &  18.39$^{+  0.08}_{-  0.08}$ &  21.02$^{+  2.44}_{-  2.18}$ &  18.96$^{+  0.30}_{-  0.20}$ &   4.36$^{+  1.07}_{-  0.68}$ &   1.57$^{+  0.25}_{-  0.16}$ &   0.026 &  5 &  5642 &  0.81 \\
12654 &  18.01$^{+  0.44}_{-  0.24}$ &  15.49$^{+  3.87}_{-  2.61}$ &  19.11$^{+  1.06}_{-  0.37}$ &   4.09$^{+  7.54}_{-  1.53}$ &   1.08$^{+  0.80}_{-  0.34}$ &   0.023 &  4 &  4041 &  0.82 \\
12732 & \multicolumn{5}{c}{NP, no fit}                                                                                                                           &   0.032 &  9 &  0749 &  0.89 \\
12754 &     NP                       &  33.94$^{+  2.26}_{-  2.50}$ &     NP                       &   6.94$^{+  0.27}_{-  0.27}$ &   0.70$^{+  0.02}_{-  0.01}$ &   0.028 &  6 &  0751 &  0.70 \\
12776 &  19.99$^{+  0.06}_{-  0.11}$ &  44.34$^{+  7.56}_{-  7.49}$ &  17.43$^{+  0.05}_{-  0.09}$ &   6.02$^{+  0.19}_{-  0.27}$ &   2.76$^{+  0.07}_{-  0.10}$ &   0.022 &  3 &  4937 &  0.88 \\
12808 & \multicolumn{5}{c}{no data}                                                                                                                              &   0.027 &  3 &  4211 &  0.93 \\
12845 &     NP                       &  17.01$^{+  2.08}_{-  2.51}$ &      NP                      &   2.43$^{+  0.30}_{-  0.35}$ &   0.71$^{+  0.13}_{-  0.12}$ &   0.020 &  7 &  4875 &  0.81 \\
\enddata
\tablecomments{Column 1: Uppsala General Catalog (UGC) galaxy identification.
Column 2 and 3: Central (uncorrected) disk surface brightness $\mu_{\rm 0,d}$ and 
disk scale-length $h$.  Column 4, 5 and 6: (Uncorrected) effective bulge surface 
brightness $\mu_{\rm e,b}$, half-light bulge radius $R_{\rm e,b}$, and bulge profile 
shape $n$. 
Column 7: Galactic extinction (via the NASA Extragalactic Database: NED).  
Column 8: Galaxy T-type.  Column 9: Heliocentric velocity (via NED).  
Column 10: Projected minor-to-major axis ratio of the outer disk (from de 
Jong 1996a).  Corrective terms for the surface brightness values are given 
in the main paper. 
}
\end{deluxetable}

\begin{deluxetable}{crrrrrr}
\tablecolumns{9}
\tablewidth{0pc}
\tablecaption{$I$-band model parameters}
\tablehead{
\colhead{Galaxy} & \colhead{$\mu_{0,d}$} & \colhead{$h$}   &  \colhead{$\mu_{e,b}$}  &
\colhead{$R_{e,b}$} & \colhead{$n$}   & \colhead{extinc} \\
\colhead{UGC \#} & \colhead{mag/$\sq\arcsec$} & \colhead{arcsec}   &  \colhead{mag/$\sq\arcsec$}  &
\colhead{arcsec} & \colhead{}   & \colhead{$I$-mag}
}
\startdata
00089 & \multicolumn{6}{c}{no fit} \\                                                                                                                           		 
00093 &  20.20$^{+  0.02}_{-  0.03}$ &  17.17$^{+  1.51}_{-  1.55}$ &  20.56$^{+  0.09}_{-  0.10}$ &   1.91$^{+  0.15}_{-  0.16}$ &   0.93$^{+  0.08}_{-  0.09}$ &  0.110 \\
00242 &  19.21$^{+  0.07}_{-  0.10}$ &  12.49$^{+  1.04}_{-  1.21}$ &  19.37$^{+  0.06}_{-  0.14}$ &   1.11$^{+  0.09}_{-  0.13}$ &   0.47$^{+  0.06}_{-  0.15}$ &  0.129 \\
00334 &  22.02$^{+  0.21}_{-  0.43}$ &  19.70$^{+  6.95}_{-  6.84}$ &  23.15$^{+  0.23}_{-  0.58}$ &   4.96$^{+  2.20}_{-  2.79}$ &   1.09$^{+  0.19}_{-  0.60}$ &  0.108 \\
00438 &  18.71$^{+  0.37}_{-  0.34}$ &  14.04$^{+  1.41}_{-  1.63}$ &  19.60$^{+  0.56}_{-  0.72}$ &   6.84$^{+  4.36}_{-  3.01}$ &   2.26$^{+  0.36}_{-  0.54}$ &  0.069 \\
00463 &  18.72$^{+  0.03}_{-  0.03}$ &  11.85$^{+  0.18}_{-  0.19}$ &  18.69$^{+  0.00}_{-  0.00}$ &   1.80$^{+  0.02}_{-  0.02}$ &   0.30$^{+  0.01}_{-  0.01}$ &  0.176 \\
00490 &  18.86$^{+  0.05}_{-  0.04}$ &  13.35$^{+  0.38}_{-  0.36}$ &  19.87$^{+  0.15}_{-  0.14}$ &   4.12$^{+  0.51}_{-  0.42}$ &   1.71$^{+  0.12}_{-  0.11}$ &  0.097 \\
00508 &  19.33$^{+  0.10}_{-  0.12}$ &  22.18$^{+  1.51}_{-  1.66}$ &  17.84$^{+  0.08}_{-  0.09}$ &   3.77$^{+  0.19}_{-  0.22}$ &   1.40$^{+  0.07}_{-  0.08}$ &  0.134 \\
00628 &  21.15$^{+  0.20}_{-  0.27}$ &  14.73$^{+  4.28}_{-  3.71}$ &  22.31$^{+  0.21}_{-  0.34}$ &   4.04$^{+  1.62}_{-  1.51}$ &   0.95$^{+  0.21}_{-  0.42}$ &  0.086 \\
01305 &  19.97$^{+  0.14}_{-  0.15}$ &  33.79$^{+  4.01}_{-  3.78}$ &  20.21$^{+  0.31}_{-  0.31}$ &  10.33$^{+  2.44}_{-  2.03}$ &   2.42$^{+  0.27}_{-  0.29}$ &  0.138 \\
01455 &  19.89$^{+  0.31}_{-  0.33}$ &  19.52$^{+  3.60}_{-  3.26}$ &  18.84$^{+  0.62}_{-  0.56}$ &   3.94$^{+  1.68}_{-  1.04}$ &   2.24$^{+  0.66}_{-  0.61}$ &  0.200 \\
01551 &  20.76$^{+  0.05}_{-  0.09}$ &  24.45$^{+  2.78}_{-  3.14}$ &  21.78$^{+  0.08}_{-  0.06}$ &   2.79$^{+  0.41}_{-  0.51}$ &   0.80$^{+  0.07}_{-  0.06}$ &  0.176 \\
01559 &  21.39$^{+  0.05}_{-  0.15}$ &  25.44$^{+  5.88}_{-  6.03}$ &  21.55$^{+  0.03}_{-  0.02}$ &   5.35$^{+  0.33}_{-  0.52}$ &   0.70$^{+  0.04}_{-  0.07}$ &  0.119 \\
01577 &  19.56$^{+  0.20}_{-  0.14}$ &  15.17$^{+  1.72}_{-  1.33}$ &  18.99$^{+  0.45}_{-  0.15}$ &   3.36$^{+  0.95}_{-  0.32}$ &   1.20$^{+  0.65}_{-  0.21}$ &  0.111 \\
01719 &  20.27$^{+  0.59}_{-  0.51}$ &  18.69$^{+  4.50}_{-  4.22}$ &  20.38$^{+  1.53}_{-  1.07}$ &   5.92$^{+ 10.52}_{-  2.99}$ &   2.46$^{+  1.32}_{-  0.96}$ &  0.158 \\
01792 &  19.45$^{+  0.07}_{-  0.10}$ &  13.82$^{+  0.91}_{-  1.05}$ &  19.10$^{+  0.05}_{-  0.05}$ &   2.10$^{+  0.11}_{-  0.13}$ &   0.60$^{+  0.05}_{-  0.06}$ &  0.173 \\
02064 &  20.11$^{+  0.11}_{-  0.21}$ &  19.74$^{+  2.50}_{-  3.14}$ &  19.83$^{+  0.20}_{-  0.31}$ &   2.70$^{+  0.43}_{-  0.59}$ &   1.27$^{+  0.18}_{-  0.19}$ &  0.256 \\
02081 &  20.77$^{+  0.14}_{-  0.25}$ &  20.65$^{+  4.20}_{-  4.76}$ &  21.57$^{+  0.46}_{-  0.23}$ &   3.37$^{+  1.63}_{-  1.11}$ &   0.81$^{+  0.38}_{-  0.26}$ &  0.050 \\
02124 &  19.90$^{+  0.11}_{-  0.15}$ &  22.39$^{+  2.03}_{-  2.21}$ &  18.31$^{+  0.03}_{-  0.03}$ &   4.61$^{+  0.07}_{-  0.10}$ &   0.78$^{+  0.09}_{-  0.08}$ &  0.062 \\
02125 &  20.66$^{+  0.46}_{-  0.84}$ &  18.17$^{+  6.33}_{-  6.95}$ &  19.50$^{+  0.92}_{-  0.82}$ &   3.41$^{+  2.25}_{-  1.46}$ &   1.82$^{+  1.39}_{-  0.30}$ &  0.274 \\
02197 &  20.16$^{+  0.04}_{-  0.07}$ &  13.40$^{+  0.61}_{-  0.80}$ &  20.77$^{+  0.10}_{-  0.16}$ &   2.86$^{+  0.23}_{-  0.32}$ &   0.84$^{+  0.10}_{-  0.16}$ &  0.417 \\
02368 &  19.68$^{+  0.05}_{-  0.06}$ &  14.36$^{+  0.64}_{-  0.65}$ &  18.99$^{+  0.09}_{-  0.11}$ &   2.41$^{+  0.14}_{-  0.15}$ &   1.46$^{+  0.08}_{-  0.07}$ &  0.324 \\
02595 &  19.72$^{+  0.12}_{-  0.33}$ &  16.91$^{+  2.01}_{-  3.31}$ &  18.21$^{+  0.15}_{-  0.36}$ &   1.67$^{+  0.17}_{-  0.37}$ &   1.53$^{+  0.12}_{-  0.20}$ &  0.334 \\
03066 &  19.27$^{+  0.15}_{-  0.34}$ &  11.00$^{+  1.18}_{-  1.93}$ &  19.73$^{+  0.08}_{-  0.22}$ &   1.63$^{+  0.30}_{-  0.58}$ &   0.56$^{+  0.09}_{-  0.30}$ &  0.599 \\
03080 &  20.01$^{+  0.09}_{-  0.20}$ &  15.84$^{+  1.88}_{-  2.62}$ &  20.36$^{+  0.09}_{-  0.16}$ &   1.45$^{+  0.20}_{-  0.36}$ &   0.55$^{+  0.09}_{-  0.16}$ &  0.167 \\
03140 &  19.51$^{+  0.14}_{-  0.18}$ &  13.14$^{+  0.12}_{-  0.48}$ &  20.21$^{+  0.51}_{-  0.55}$ &   7.18$^{+  2.85}_{-  2.15}$ &   2.83$^{+  0.46}_{-  0.51}$ &  0.157 \\
04126 &  19.69$^{+  0.15}_{-  0.23}$ &  19.15$^{+  2.04}_{-  2.58}$ &  18.66$^{+  0.23}_{-  0.27}$ &   2.69$^{+  0.38}_{-  0.45}$ &   1.48$^{+  0.36}_{-  0.23}$ &  0.101 \\
04256 &     NP                       &  14.21$^{+  0.84}_{-  0.86}$ &     NP                       &   1.96$^{+  0.50}_{-  0.34}$ &   2.08$^{+  0.44}_{-  0.38}$ &  0.104 \\
04308 &  19.59$^{+  0.04}_{-  0.06}$ &  19.57$^{+  1.01}_{-  1.09}$ &  19.16$^{+  0.10}_{-  0.10}$ &   2.01$^{+  0.13}_{-  0.13}$ &   1.09$^{+  0.10}_{-  0.08}$ &  0.076 \\
04368 &  19.71$^{+  0.03}_{-  0.04}$ &  17.15$^{+  0.64}_{-  0.67}$ &  20.13$^{+  0.03}_{-  0.02}$ &   2.71$^{+  0.10}_{-  0.11}$ &   0.66$^{+  0.03}_{-  0.03}$ &  0.073 \\
04375 & \multicolumn{6}{c}{no data} \\
04422 &  20.11$^{+  0.06}_{-  0.13}$ &  28.32$^{+  4.06}_{-  4.70}$ &  18.47$^{+  0.04}_{-  0.06}$ &   3.43$^{+  0.09}_{-  0.14}$ &   0.77$^{+  0.04}_{-  0.06}$ &  0.077 \\
04458 &  21.34$^{+  0.17}_{-  0.22}$ &  32.83$^{+  3.95}_{-  3.81}$ &  19.17$^{+  0.10}_{-  0.14}$ &   8.02$^{+  0.49}_{-  0.62}$ &   3.04$^{+  0.12}_{-  0.16}$ &  0.067 \\
05103 & \multicolumn{6}{c}{no data} \\
05303 &  19.40$^{+  0.02}_{-  0.03}$ &  31.24$^{+  0.78}_{-  0.82}$ &  19.34$^{+  0.01}_{-  0.01}$ &   4.22$^{+  0.08}_{-  0.09}$ &   0.67$^{+  0.01}_{-  0.01}$ &  0.066 \\
05510 &  19.07$^{+  0.04}_{-  0.04}$ &  18.55$^{+  0.52}_{-  0.56}$ &  18.61$^{+  0.11}_{-  0.10}$ &   2.28$^{+  0.15}_{-  0.14}$ &   1.36$^{+  0.11}_{-  0.10}$ &  0.045 \\
05554 &  18.82$^{+  0.12}_{-  0.19}$ &  16.17$^{+  1.37}_{-  1.72}$ &  17.71$^{+  0.15}_{-  0.23}$ &   2.31$^{+  0.23}_{-  0.30}$ &   1.29$^{+  0.14}_{-  0.22}$ &  0.052 \\
05633 &      NP                      &  26.50$^{+  4.18}_{-  3.66}$ &      NP                      &   8.65$^{+  2.79}_{-  2.85}$ &   1.12$^{+  0.11}_{-  0.27}$  & 0.084 \\
05842 &  20.06$^{+  0.00}_{-  0.00}$ &  37.17$^{+  0.46}_{-  0.47}$ &  20.29$^{+  0.00}_{-  0.01}$ &   3.12$^{+  0.01}_{-  0.01}$ &   0.73$^{+  0.00}_{-  0.01}$ &  0.054 \\
06028 &  \multicolumn{6}{c}{no fit} \\ 
06077 &  19.20$^{+  0.06}_{-  0.08}$ &  17.67$^{+  0.96}_{-  1.13}$ &  18.97$^{+  0.04}_{-  0.06}$ &   2.20$^{+  0.12}_{-  0.16}$ &   0.76$^{+  0.03}_{-  0.04}$ &  0.041 \\
06123 & \multicolumn{6}{c}{no data} \\
06277 &  19.15$^{+  0.05}_{-  0.05}$ &  25.74$^{+  1.02}_{-  1.00}$ &  19.46$^{+  0.20}_{-  0.16}$ &   4.86$^{+  0.66}_{-  0.50}$ &   2.41$^{+  0.18}_{-  0.14}$ &  0.043 \\
06445 &  18.48$^{+  0.02}_{-  0.02}$ &  13.97$^{+  0.20}_{-  0.20}$ &  17.94$^{+  0.01}_{-  0.01}$ &   3.55$^{+  0.03}_{-  0.03}$ &   0.88$^{+  0.01}_{-  0.01}$ &  0.051 \\
06453 &  18.79$^{+  0.06}_{-  0.06}$ &  15.79$^{+  0.36}_{-  0.35}$ &  20.35$^{+  0.12}_{-  0.12}$ &   7.68$^{+  1.09}_{-  0.94}$ &   1.67$^{+  0.08}_{-  0.09}$ &  0.050 \\
06460 & \multicolumn{6}{c}{no data} \\
06536 & \multicolumn{6}{c}{no data} \\
06693 &  19.85$^{+  0.05}_{-  0.06}$ &  18.39$^{+  1.02}_{-  1.15}$ &  19.68$^{+  0.04}_{-  0.05}$ &   1.70$^{+  0.08}_{-  0.10}$ &   0.70$^{+  0.03}_{-  0.04}$ &  0.047 \\
06746 & \multicolumn{6}{c}{no data} \\
06754 & \multicolumn{6}{c}{no data} \\
07169 &       NP                     &   9.82$^{+  0.36}_{-  0.36}$ &       NP                     &   2.31$^{+  0.20}_{-  0.18}$ &   1.07$^{+  0.09}_{-  0.09}$ &  0.065 \\
07315 &  17.98$^{+  0.03}_{-  0.03}$ &  14.53$^{+  0.34}_{-  0.35}$ &  18.66$^{+  0.05}_{-  0.05}$ &   2.26$^{+  0.12}_{-  0.12}$ &   0.67$^{+  0.05}_{-  0.05}$ &  0.058 \\
07450 &  19.24$^{+  0.07}_{-  0.09}$ &  55.46$^{+  3.51}_{-  4.00}$ &  18.33$^{+  0.02}_{-  0.03}$ &   8.00$^{+  0.18}_{-  0.22}$ &   0.37$^{+  0.03}_{-  0.04}$ &  0.051 \\
07523 &  19.25$^{+  0.10}_{-  0.12}$ &  30.54$^{+  2.26}_{-  2.37}$ &  17.95$^{+  0.10}_{-  0.12}$ &   5.68$^{+  0.37}_{-  0.40}$ &   1.68$^{+  0.09}_{-  0.10}$ &  0.059 \\
07594 & \multicolumn{6}{c}{no data} \\
07876 &        NP                    &  19.90$^{+  1.00}_{-  1.12}$ &      NP                      &   2.34$^{+  0.30}_{-  0.35}$ &   0.85$^{+  0.03}_{-  0.02}$ &  0.053 \\
07901 &  17.61$^{+  0.00}_{-  0.00}$ &  16.73$^{+  0.07}_{-  0.07}$ &  17.64$^{+  0.01}_{-  0.01}$ &   2.82$^{+  0.01}_{-  0.01}$ &   1.16$^{+  0.01}_{-  0.01}$ &  0.052 \\
08279 & \multicolumn{6}{c}{no fit} \\
08289 &  19.78$^{+  0.09}_{-  0.16}$ &  24.63$^{+  2.54}_{-  3.09}$ &  17.56$^{+  0.07}_{-  0.09}$ &   2.80$^{+  0.08}_{-  0.12}$ &   1.00$^{+  0.08}_{-  0.10}$ &  0.046 \\
08865 &  20.27$^{+  0.02}_{-  0.03}$ &  29.71$^{+  0.48}_{-  0.48}$ &  18.94$^{+  0.02}_{-  0.02}$ &   6.51$^{+  0.08}_{-  0.08}$ &   1.70$^{+  0.02}_{-  0.02}$ &  0.040 \\
09024 &  23.16$^{+  0.00}_{-  0.00}$ &  31.61$^{+  0.00}_{-  0.00}$ &  20.83$^{+  0.00}_{-  0.00}$ &   3.69$^{+  0.00}_{-  0.00}$ &   1.26$^{+  0.00}_{-  0.00}$ &  0.053 \\
09061 &  20.80$^{+  0.02}_{-  0.03}$ &  51.86$^{+  5.02}_{-  5.18}$ &  18.54$^{+  0.02}_{-  0.03}$ &   3.98$^{+  0.04}_{-  0.05}$ &   1.16$^{+  0.04}_{-  0.04}$ &  0.046 \\
09481 &  19.46$^{+  0.14}_{-  0.37}$ &  15.84$^{+  1.85}_{-  3.16}$ &  19.30$^{+  0.25}_{-  0.32}$ &   2.25$^{+  0.46}_{-  0.69}$ &   0.94$^{+  0.23}_{-  0.23}$ &  0.033 \\
09915 &  19.25$^{+  0.05}_{-  0.05}$ &  17.42$^{+  0.60}_{-  0.62}$ &  19.00$^{+  0.06}_{-  0.06}$ &   3.30$^{+  0.16}_{-  0.16}$ &   1.32$^{+  0.05}_{-  0.04}$ &  0.083 \\
09926 & \multicolumn{6}{c}{no data} \\
09943 &  18.80$^{+  0.22}_{-  0.21}$ &  20.87$^{+  2.00}_{-  1.80}$ &  19.27$^{+  0.15}_{-  0.15}$ &   8.85$^{+  1.69}_{-  1.50}$ &   1.42$^{+  0.11}_{-  0.11}$ &  0.081 \\
10083 &  18.82$^{+  0.04}_{-  0.04}$ &  14.77$^{+  0.38}_{-  0.39}$ &  20.85$^{+  0.04}_{-  0.07}$ &   4.09$^{+  0.46}_{-  0.50}$ &   1.55$^{+  0.00}_{-  0.01}$ &  0.071 \\
10437 &  22.41$^{+  0.05}_{-  0.17}$ &  31.51$^{+  8.57}_{-  8.35}$ &  21.52$^{+  0.02}_{-  0.03}$ &   7.54$^{+  0.20}_{-  0.34}$ &   0.88$^{+  0.02}_{-  0.04}$ &  0.018 \\
10445 &  20.62$^{+  0.05}_{-  0.07}$ &  21.23$^{+  1.84}_{-  1.85}$ &  21.31$^{+  0.00}_{-  0.00}$ &   5.47$^{+  0.25}_{-  0.29}$ &   0.68$^{+  0.03}_{-  0.03}$ &  0.062 \\
10584 &  20.00$^{+  0.03}_{-  0.04}$ &  22.12$^{+  1.58}_{-  1.66}$ &  19.14$^{+  0.03}_{-  0.04}$ &   2.10$^{+  0.05}_{-  0.06}$ &   0.77$^{+  0.03}_{-  0.04}$ &  0.025 \\
11628 &  20.23$^{+  0.31}_{-  0.52}$ &  30.00$^{+  7.16}_{-  7.40}$ &  18.86$^{+  0.24}_{-  0.41}$ &   7.05$^{+  1.18}_{-  1.66}$ &   2.17$^{+  0.22}_{-  0.36}$ &  0.190 \\
11708 &  19.28$^{+  0.08}_{-  0.12}$ &  14.49$^{+  1.54}_{-  1.76}$ &  19.07$^{+  0.32}_{-  0.14}$ &   2.18$^{+  0.40}_{-  0.26}$ &   1.12$^{+  0.35}_{-  0.12}$ &  0.131 \\
11872 &  18.30$^{+  1.16}_{-  0.78}$ &  17.31$^{+  4.49}_{-  3.34}$ &  18.18$^{+  0.42}_{-  0.52}$ &   9.20$^{+  4.77}_{-  3.72}$ &   1.53$^{+  0.31}_{-  0.41}$ &  0.139 \\
12151 &  21.47$^{+  0.10}_{-  0.14}$ &  19.88$^{+  4.32}_{-  4.00}$ &  22.47$^{+  0.20}_{-  0.05}$ &   4.23$^{+  1.08}_{-  0.74}$ &   0.69$^{+  0.22}_{-  0.14}$ &  0.135 \\
12343 &  19.35$^{+  0.08}_{-  0.08}$ &  37.28$^{+  2.97}_{-  2.82}$ &  19.12$^{+  0.12}_{-  0.06}$ &   7.41$^{+  0.64}_{-  0.46}$ &   1.21$^{+  0.12}_{-  0.05}$ &  0.217 \\
12379 &  19.86$^{+  0.15}_{-  0.16}$ &  19.79$^{+  1.81}_{-  1.73}$ &  18.98$^{+  0.30}_{-  0.28}$ &   4.18$^{+  0.81}_{-  0.63}$ &   2.30$^{+  0.32}_{-  0.30}$ &  0.163 \\
12391 &  19.72$^{+  0.07}_{-  0.12}$ &  15.39$^{+  1.60}_{-  1.86}$ &  18.92$^{+  0.22}_{-  0.29}$ &   0.95$^{+  0.13}_{-  0.16}$ &   1.23$^{+  0.21}_{-  0.20}$ &  0.167 \\
12511 &  19.97$^{+  0.02}_{-  0.03}$ &  14.94$^{+  0.90}_{-  0.90}$ &  20.03$^{+  0.07}_{-  0.07}$ &   2.00$^{+  0.10}_{-  0.10}$ &   0.89$^{+  0.07}_{-  0.07}$ &  0.077 \\
12614 &  19.17$^{+  0.03}_{-  0.04}$ &  21.04$^{+  0.75}_{-  0.80}$ &  17.99$^{+  0.04}_{-  0.05}$ &   1.70$^{+  0.05}_{-  0.05}$ &   1.22$^{+  0.04}_{-  0.04}$ &  0.095 \\
12638 &  20.43$^{+  0.04}_{-  0.05}$ &  22.88$^{+  1.84}_{-  1.79}$ &  21.02$^{+  0.23}_{-  0.22}$ &   3.85$^{+  0.66}_{-  0.54}$ &   2.10$^{+  0.20}_{-  0.19}$ &  0.136 \\
12654 &  19.79$^{+  0.06}_{-  0.06}$ &  17.01$^{+  0.83}_{-  0.81}$ &  20.85$^{+  0.18}_{-  0.14}$ &   3.71$^{+  0.64}_{-  0.48}$ &   1.36$^{+  0.14}_{-  0.11}$ &  0.121 \\
12732 & \multicolumn{6}{c}{no fit} \\
12754 &  19.85$^{+  0.08}_{-  0.13}$ &  37.30$^{+  4.13}_{-  4.64}$ &  20.41$^{+  0.01}_{-  0.00}$ &   6.67$^{+  0.52}_{-  0.75}$ &   0.69$^{+  0.03}_{-  0.04}$ &  0.145 \\
12776 &  21.19$^{+  0.12}_{-  0.24}$ &  37.22$^{+  8.44}_{-  8.39}$ &  19.07$^{+  0.12}_{-  0.18}$ &   5.36$^{+  0.36}_{-  0.51}$ &   2.15$^{+  0.14}_{-  0.19}$ &  0.118 \\
12808 &  18.33$^{+  0.05}_{-  0.04}$ &  12.64$^{+  0.33}_{-  0.33}$ &  18.42$^{+  0.25}_{-  0.03}$ &   2.99$^{+  0.45}_{-  0.08}$ &   3.22$^{+  0.41}_{-  0.01}$ &  0.144 \\
12845 &  20.61$^{+  0.09}_{-  0.16}$ &  21.13$^{+  3.28}_{-  3.67}$ &  20.72$^{+  0.17}_{-  0.20}$ &   2.85$^{+  0.45}_{-  0.50}$ &   0.93$^{+  0.15}_{-  0.17}$ &  0.107 \\
\enddata
\tablecomments{See Table 1 for a description of each column.}
\end{deluxetable}

\begin{deluxetable}{crrrrrr}
\tablecolumns{9}
\tablewidth{0pc}
\tablecaption{$R$-band model parameters}
\tablehead{
\colhead{Galaxy} & \colhead{$\mu_{0,d}$} & \colhead{$h$}   &  \colhead{$\mu_{e,b}$}  &
\colhead{$R_{e,b}$} & \colhead{$n$}   & \colhead{extinc} \\
\colhead{UGC \#} & \colhead{mag/$\sq\arcsec$} & \colhead{arcsec}   &  \colhead{mag/$\sq\arcsec$}  &
\colhead{arcsec} & \colhead{}   & \colhead{$R$-mag}
}
\startdata
00089 & \multicolumn{6}{c}{no fit} \\                                                                                                                           	  
00093 &  20.95$^{+  0.08}_{-  0.05}$ &  18.84$^{+  2.13}_{-  1.81}$ &  21.26$^{+  0.98}_{-  0.16}$ &   1.90$^{+  1.38}_{-  0.23}$ &   1.08$^{+  0.95}_{-  0.14}$ &  0.152 \\
00242 &  19.87$^{+  0.06}_{-  0.07}$ &  12.57$^{+  0.67}_{-  0.74}$ &  19.89$^{+  0.11}_{-  0.24}$ &   0.91$^{+  0.07}_{-  0.05}$ &   0.44$^{+  0.11}_{-  0.29}$ &  0.178 \\
00334 &  22.26$^{+  0.12}_{-  0.10}$ &  20.89$^{+  3.50}_{-  2.92}$ &  23.64$^{+  0.47}_{-  0.25}$ &   4.21$^{+  2.56}_{-  1.13}$ &   1.19$^{+  0.34}_{-  0.19}$ &  0.148 \\
00438 &  19.30$^{+  0.31}_{-  0.32}$ &  14.04$^{+  1.48}_{-  1.65}$ &  20.44$^{+  0.39}_{-  0.80}$ &   6.89$^{+  3.44}_{-  3.32}$ &   2.24$^{+  0.20}_{-  0.60}$ &  0.095 \\
00463 & \multicolumn{6}{c}{no data} \\
00490 &  19.53$^{+  0.01}_{-  0.01}$ &  13.58$^{+  0.15}_{-  0.15}$ &  20.16$^{+  0.04}_{-  0.02}$ &   3.03$^{+  0.08}_{-  0.06}$ &   1.38$^{+  0.03}_{-  0.03}$ &  0.133 \\
00508 &  20.07$^{+  0.04}_{-  0.05}$ &  24.57$^{+  0.86}_{-  0.88}$ &  18.59$^{+  0.04}_{-  0.03}$ &   4.01$^{+  0.09}_{-  0.10}$ &   1.51$^{+  0.04}_{-  0.04}$ &  0.184 \\
00628 &  21.60$^{+  0.13}_{-  0.14}$ &  13.55$^{+  2.07}_{-  1.93}$ &  22.74$^{+  0.14}_{-  0.14}$ &   3.56$^{+  0.83}_{-  0.74}$ &   0.83$^{+  0.15}_{-  0.19}$ &  0.118 \\
01305 &  20.58$^{+  0.04}_{-  0.03}$ &  34.68$^{+  1.08}_{-  1.05}$ &  20.56$^{+  0.07}_{-  0.06}$ &   7.91$^{+  0.38}_{-  0.35}$ &   1.99$^{+  0.07}_{-  0.06}$ &  0.190 \\
01455 &  20.74$^{+  0.10}_{-  0.10}$ &  21.64$^{+  1.30}_{-  1.25}$ &  19.77$^{+  0.22}_{-  0.21}$ &   4.24$^{+  0.56}_{-  0.47}$ &   2.40$^{+  0.24}_{-  0.23}$ &  0.276 \\
01551 &  21.21$^{+  0.04}_{-  0.04}$ &  24.45$^{+  1.55}_{-  1.68}$ &  22.33$^{+  0.02}_{-  0.01}$ &   2.55$^{+  0.20}_{-  0.23}$ &   0.57$^{+  0.04}_{-  0.05}$ &  0.242 \\
01559 &  21.72$^{+  0.02}_{-  0.02}$ &  22.06$^{+  1.01}_{-  1.00}$ &  21.94$^{+  0.01}_{-  0.00}$ &   5.30$^{+  0.07}_{-  0.06}$ &   0.66$^{+  0.01}_{-  0.00}$ &  0.165 \\
01577 &  20.26$^{+  0.08}_{-  0.02}$ &  16.86$^{+  0.95}_{-  0.56}$ &  19.63$^{+  0.16}_{-  0.02}$ &   3.16$^{+  0.31}_{-  0.06}$ &   1.27$^{+  0.14}_{-  0.03}$ &  0.153 \\
01719 &  20.83$^{+  0.18}_{-  0.17}$ &  19.28$^{+  1.89}_{-  1.75}$ &  20.70$^{+  0.43}_{-  0.37}$ &   4.85$^{+  1.55}_{-  1.04}$ &   2.22$^{+  0.39}_{-  0.33}$ &  0.218 \\
01792 &  20.04$^{+  0.04}_{-  0.04}$ &  14.66$^{+  0.53}_{-  0.55}$ &  19.71$^{+  0.02}_{-  0.02}$ &   2.04$^{+  0.05}_{-  0.06}$ &   0.58$^{+  0.02}_{-  0.02}$ &  0.239 \\
02064 &  20.73$^{+  0.08}_{-  0.13}$ &  18.89$^{+  1.65}_{-  1.98}$ &  20.42$^{+  0.14}_{-  0.22}$ &   2.32$^{+  0.28}_{-  0.36}$ &   1.23$^{+  0.07}_{-  0.09}$ &  0.353 \\
02081 &  21.36$^{+  0.07}_{-  0.10}$ &  18.50$^{+  2.16}_{-  2.30}$ &  22.21$^{+  0.09}_{-  0.05}$ &   2.96$^{+  0.39}_{-  0.43}$ &   0.66$^{+  0.12}_{-  0.06}$ &  0.069 \\
02124 &  20.63$^{+  0.08}_{-  0.09}$ &  23.38$^{+  1.59}_{-  1.68}$ &  19.04$^{+  0.02}_{-  0.02}$ &   4.65$^{+  0.06}_{-  0.06}$ &   0.76$^{+  0.06}_{-  0.05}$ &  0.086 \\
02125 &  21.31$^{+  0.10}_{-  0.11}$ &  21.62$^{+  2.38}_{-  2.42}$ &  20.11$^{+  0.14}_{-  0.13}$ &   3.17$^{+  0.26}_{-  0.23}$ &   1.54$^{+  0.23}_{-  0.20}$ &  0.378 \\
02197 &  20.90$^{+  0.15}_{-  0.09}$ &  14.93$^{+  1.54}_{-  1.11}$ &  21.83$^{+  0.53}_{-  0.28}$ &   3.62$^{+  1.98}_{-  0.54}$ &   1.08$^{+  0.42}_{-  0.37}$ &  0.574 \\
02368 &  20.44$^{+  0.01}_{-  0.01}$ &  15.77$^{+  0.22}_{-  0.21}$ &  19.28$^{+  0.04}_{-  0.03}$ &   1.87$^{+  0.04}_{-  0.02}$ &   1.79$^{+  0.04}_{-  0.03}$ &  0.447 \\
02595 &  20.56$^{+  0.05}_{-  0.06}$ &  18.43$^{+  0.98}_{-  1.08}$ &  19.18$^{+  0.06}_{-  0.06}$ &   1.84$^{+  0.07}_{-  0.08}$ &   1.15$^{+  0.04}_{-  0.03}$ &  0.461 \\
03066 &  20.18$^{+  0.07}_{-  0.10}$ &  12.26$^{+  0.78}_{-  0.91}$ &  20.68$^{+  0.04}_{-  0.05}$ &   1.66$^{+  0.16}_{-  0.20}$ &   0.58$^{+  0.04}_{-  0.06}$ &  0.825 \\
03080 &  20.65$^{+  0.05}_{-  0.05}$ &  16.23$^{+  0.88}_{-  0.95}$ &  21.01$^{+  0.06}_{-  0.08}$ &   1.43$^{+  0.09}_{-  0.10}$ &   0.34$^{+  0.06}_{-  0.09}$ &  0.231 \\
03140 & \multicolumn{6}{c}{no data} \\
04126 &  20.24$^{+  0.11}_{-  0.06}$ &  18.41$^{+  1.40}_{-  0.89}$ &  19.24$^{+  0.13}_{-  0.04}$ &   2.41$^{+  0.21}_{-  0.06}$ &   1.30$^{+  0.11}_{-  0.12}$ &  0.139 \\
04256 &     NP                       &  14.70$^{+  0.63}_{-  0.65}$ &      NP                      &   1.38$^{+  0.16}_{-  0.13}$ &   1.68$^{+  0.29}_{-  0.23}$ &  0.143 \\
04308 &  20.05$^{+  0.06}_{-  0.05}$ &  18.06$^{+  0.85}_{-  0.95}$ &  19.50$^{+  0.07}_{-  0.07}$ &   1.66$^{+  0.09}_{-  0.09}$ &   0.76$^{+  0.06}_{-  0.06}$ &  0.104 \\
04368 &  20.29$^{+  0.07}_{-  0.08}$ &  16.41$^{+  1.00}_{-  1.14}$ &  20.78$^{+  0.03}_{-  0.01}$ &   2.51$^{+  0.17}_{-  0.22}$ &   0.52$^{+  0.05}_{-  0.04}$ &  0.100 \\
04375 &  19.91$^{+  0.07}_{-  0.07}$ &  18.14$^{+  0.84}_{-  0.89}$ &  20.91$^{+  0.18}_{-  0.26}$ &   3.24$^{+  0.62}_{-  0.60}$ &   1.22$^{+  0.11}_{-  0.22}$ &  0.120 \\
04422 &  20.72$^{+  0.03}_{-  0.04}$ &  29.83$^{+  2.21}_{-  2.34}$ &  19.11$^{+  0.02}_{-  0.01}$ &   3.49$^{+  0.04}_{-  0.05}$ &   0.67$^{+  0.02}_{-  0.02}$ &  0.107 \\
04458 &  21.32$^{+  0.40}_{-  0.59}$ &  26.62$^{+  6.45}_{-  6.05}$ &  19.45$^{+  0.35}_{-  0.53}$ &   6.53$^{+  1.41}_{-  1.68}$ &   2.88$^{+  0.40}_{-  0.61}$ &  0.093 \\
05103 &  19.35$^{+  0.07}_{-  0.10}$ &  15.22$^{+  0.69}_{-  0.85}$ &  18.71$^{+  0.06}_{-  0.07}$ &   1.97$^{+  0.10}_{-  0.12}$ &   0.78$^{+  0.05}_{-  0.06}$ &  0.071 \\
05303 &  20.01$^{+  0.04}_{-  0.04}$ &  32.27$^{+  1.38}_{-  1.49}$ &  19.81$^{+  0.03}_{-  0.03}$ &   3.71$^{+  0.13}_{-  0.14}$ &   0.91$^{+  0.03}_{-  0.01}$ &  0.091 \\
05510 &  19.54$^{+  0.04}_{-  0.02}$ &  18.66$^{+  0.41}_{-  0.43}$ &  18.94$^{+  0.08}_{-  0.06}$ &   2.03$^{+  0.09}_{-  0.08}$ &   1.34$^{+  0.07}_{-  0.07}$ &  0.062 \\
05554 &  19.54$^{+  0.03}_{-  0.03}$ &  17.33$^{+  0.40}_{-  0.40}$ &  18.30$^{+  0.05}_{-  0.04}$ &   2.19$^{+  0.06}_{-  0.06}$ &   1.36$^{+  0.05}_{-  0.04}$ &  0.072 \\
05633 &  21.86$^{+  0.15}_{-  0.12}$ &  28.26$^{+  5.34}_{-  4.24}$ &  22.68$^{+  0.44}_{-  0.24}$ &   5.51$^{+  2.65}_{-  1.12}$ &   0.79$^{+  0.37}_{-  0.24}$ &  0.115 \\
05842 &  20.62$^{+  0.00}_{-  0.00}$ &  37.12$^{+  0.64}_{-  0.63}$ &  20.45$^{+  0.01}_{-  0.01}$ &   2.36$^{+  0.03}_{-  0.01}$ &   0.55$^{+  0.02}_{-  0.00}$ &  0.075 \\
06028 &  \multicolumn{6}{c}{no fit} \\
06077 &  19.71$^{+  0.04}_{-  0.03}$ &  17.31$^{+  0.49}_{-  0.52}$ &  19.46$^{+  0.03}_{-  0.02}$ &   1.99$^{+  0.06}_{-  0.06}$ &   0.78$^{+  0.02}_{-  0.01}$ &  0.057 \\
06123 &     NP                       &  25.71$^{+  1.03}_{-  1.08}$ &       NP                     &   3.10$^{+  0.16}_{-  0.16}$ &   1.31$^{+  0.07}_{-  0.06}$ &  0.065 \\
06277 &  19.67$^{+  0.05}_{-  0.04}$ &  24.68$^{+  0.89}_{-  0.86}$ &  19.91$^{+  0.29}_{-  0.23}$ &   4.05$^{+  0.75}_{-  0.53}$ &   2.59$^{+  0.29}_{-  0.23}$ &  0.059 \\
06445 &  19.16$^{+  0.02}_{-  0.02}$ &  14.38$^{+  0.16}_{-  0.15}$ &  18.69$^{+  0.02}_{-  0.01}$ &   3.60$^{+  0.05}_{-  0.03}$ &   0.97$^{+  0.02}_{-  0.01}$ &  0.070 \\
06453 &      NP                      &  17.52$^{+  1.86}_{-  1.76}$ &      NP                      &   8.96$^{+  8.05}_{-  4.15}$ &   1.83$^{+  0.37}_{-  0.46}$ &  0.069 \\
06460 &  19.63$^{+  0.01}_{-  0.00}$ &  27.79$^{+  0.26}_{-  0.26}$ &  19.07$^{+  0.03}_{-  0.00}$ &   2.54$^{+  0.03}_{-  0.01}$ &   1.14$^{+  0.03}_{-  0.01}$ &  0.065 \\
06536 &      NP                      &  24.68$^{+  2.70}_{-  2.38}$ &       NP                     &   5.80$^{+  0.33}_{-  0.25}$ &   2.38$^{+  0.11}_{-  0.04}$ &  0.062 \\
06693 &  20.35$^{+  0.06}_{-  0.08}$ &  16.94$^{+  0.95}_{-  1.13}$ &  20.32$^{+  0.05}_{-  0.06}$ &   1.39$^{+  0.10}_{-  0.12}$ &   0.69$^{+  0.04}_{-  0.04}$ &  0.065 \\
06746 &  20.11$^{+  0.06}_{-  0.05}$ &  17.97$^{+  0.56}_{-  0.49}$ &  19.08$^{+  0.13}_{-  0.07}$ &   3.43$^{+  0.22}_{-  0.13}$ &   1.54$^{+  0.22}_{-  0.14}$ &  0.065 \\
06754 &  21.35$^{+  0.07}_{-  0.07}$ &  30.13$^{+  2.15}_{-  2.06}$ &  20.27$^{+  0.20}_{-  0.18}$ &   4.34$^{+  0.47}_{-  0.37}$ &   2.35$^{+  0.32}_{-  0.27}$ &  0.072 \\
07169 &  18.62$^{+  0.04}_{-  0.02}$ &  10.58$^{+  0.25}_{-  0.22}$ &  18.88$^{+  0.12}_{-  0.03}$ &   2.11$^{+  0.15}_{-  0.08}$ &   1.07$^{+  0.12}_{-  0.02}$ &  0.090 \\
07315 &  18.59$^{+  0.05}_{-  0.04}$ &  14.37$^{+  0.42}_{-  0.45}$ &  19.16$^{+  0.10}_{-  0.11}$ &   2.01$^{+  0.17}_{-  0.21}$ &   0.86$^{+  0.09}_{-  0.09}$ &  0.080 \\
07450 &  19.81$^{+  0.07}_{-  0.07}$ &  58.78$^{+  3.44}_{-  3.84}$ &  18.62$^{+  0.03}_{-  0.02}$ &   7.51$^{+  0.12}_{-  0.14}$ &   0.38$^{+  0.03}_{-  0.03}$ &  0.070 \\
07523 &  19.84$^{+  0.06}_{-  0.06}$ &  29.33$^{+  1.23}_{-  1.26}$ &  18.56$^{+  0.05}_{-  0.06}$ &   5.58$^{+  0.19}_{-  0.20}$ &   1.46$^{+  0.05}_{-  0.04}$ &  0.081 \\
07594 &  19.92$^{+  0.26}_{-  0.17}$ &  56.61$^{+  3.76}_{-  4.39}$ &  20.41$^{+  1.01}_{-  0.35}$ &  23.29$^{+ 20.60}_{-  5.02}$ &   3.46$^{+  0.95}_{-  0.25}$ &  0.074 \\
07876 &      NP                      &  20.70$^{+  0.56}_{-  0.57}$ &       NP                     &   2.35$^{+  0.14}_{-  0.13}$ &   0.83$^{+  0.01}_{-  0.01}$ &  0.073 \\
07901 &  18.23$^{+  0.01}_{-  0.00}$ &  16.95$^{+  0.06}_{-  0.05}$ &  18.26$^{+  0.01}_{-  0.00}$ &   2.58$^{+  0.01}_{-  0.00}$ &   1.17$^{+  0.01}_{-  0.00}$ &  0.071 \\
08279 & \multicolumn{6}{c}{no fit} \\
08289 &  20.29$^{+  0.06}_{-  0.09}$ &  23.96$^{+  1.44}_{-  1.78}$ &  17.94$^{+  0.03}_{-  0.03}$ &   2.66$^{+  0.03}_{-  0.05}$ &   0.79$^{+  0.03}_{-  0.04}$ &  0.064 \\
08865 &  20.60$^{+  0.06}_{-  0.06}$ &  29.15$^{+  1.18}_{-  1.20}$ &  19.38$^{+  0.05}_{-  0.05}$ &   5.81$^{+  0.18}_{-  0.19}$ &   1.53$^{+  0.05}_{-  0.04}$ &  0.055 \\
09024 &      NP                      &  32.37$^{+  5.45}_{-  5.92}$ &       NP                     &   5.06$^{+  0.14}_{-  0.22}$ &   0.84$^{+  0.05}_{-  0.07}$ &  0.073 \\
09061 &  21.25$^{+  0.02}_{-  0.01}$ &  50.00$^{+  2.03}_{-  2.08}$ &  19.12$^{+  0.01}_{-  0.01}$ &   3.77$^{+  0.02}_{-  0.03}$ &   1.09$^{+  0.02}_{-  0.02}$ &  0.064 \\
09481 &  20.06$^{+  0.09}_{-  0.11}$ &  16.82$^{+  1.29}_{-  1.52}$ &  19.98$^{+  0.09}_{-  0.09}$ &   2.31$^{+  0.20}_{-  0.23}$ &   0.82$^{+  0.08}_{-  0.11}$ &  0.046 \\
09915 &      NP                      &  16.11$^{+  0.93}_{-  1.09}$ &      NP                      &   2.79$^{+  0.22}_{-  0.27}$ &   1.18$^{+  0.05}_{-  0.05}$ &  0.115 \\
09926 &      NP                      &  18.90$^{+  0.74}_{-  0.75}$ &      NP                      &   6.71$^{+  1.86}_{-  1.40}$ &   2.22$^{+  0.27}_{-  0.28}$ &  0.145 \\
09943 &  19.18$^{+  0.12}_{-  0.11}$ &  19.62$^{+  1.03}_{-  1.01}$ &  19.74$^{+  0.09}_{-  0.09}$ &   7.05$^{+  0.80}_{-  0.77}$ &   1.26$^{+  0.07}_{-  0.07}$ &  0.111 \\
10083 &  19.44$^{+  0.04}_{-  0.04}$ &  15.72$^{+  0.45}_{-  0.43}$ &  21.54$^{+  0.04}_{-  0.04}$ &   4.56$^{+  0.55}_{-  0.53}$ &   1.52$^{+  0.01}_{-  0.00}$ &  0.097 \\
10437 &  23.60$^{+  0.00}_{-  0.22}$ &  56.10$^{+ 23.99}_{- 30.53}$ &  21.87$^{+  0.02}_{-  0.05}$ &   8.27$^{+  0.07}_{-  0.41}$ &   0.85$^{+  0.02}_{-  0.06}$ &  0.025 \\
10445 &  21.00$^{+  0.09}_{-  0.18}$ &  20.42$^{+  3.51}_{-  3.89}$ &  21.76$^{+  0.10}_{-  0.02}$ &   4.50$^{+  0.35}_{-  0.61}$ &   0.55$^{+  0.04}_{-  0.04}$ &  0.086 \\
10584 &  20.57$^{+  0.03}_{-  0.02}$ &  20.72$^{+  0.96}_{-  0.99}$ &  19.82$^{+  0.02}_{-  0.02}$ &   2.06$^{+  0.04}_{-  0.03}$ &   0.58$^{+  0.03}_{-  0.02}$ &  0.034 \\
11628 &  20.73$^{+  0.12}_{-  0.15}$ &  33.62$^{+  3.42}_{-  3.46}$ &  19.43$^{+  0.11}_{-  0.14}$ &   7.04$^{+  0.52}_{-  0.61}$ &   2.22$^{+  0.11}_{-  0.12}$ &  0.262 \\
11708 &  20.15$^{+  0.18}_{-  0.17}$ &  17.80$^{+  3.02}_{-  2.79}$ &  20.24$^{+  0.86}_{-  0.40}$ &   2.84$^{+  2.06}_{-  0.69}$ &   1.47$^{+  0.77}_{-  0.37}$ &  0.180 \\
11872 &  19.07$^{+  0.45}_{-  0.32}$ &  17.75$^{+  1.70}_{-  1.39}$ &  19.09$^{+  0.19}_{-  0.18}$ &  10.07$^{+  2.24}_{-  1.77}$ &   1.53$^{+  0.15}_{-  0.14}$ &  0.192 \\
12151 &  22.11$^{+  0.18}_{-  0.33}$ &  20.78$^{+  8.09}_{-  6.93}$ &  23.25$^{+  0.30}_{-  0.37}$ &   5.18$^{+  2.56}_{-  2.33}$ &   1.02$^{+  0.27}_{-  0.46}$ &  0.186 \\
12343 &  19.97$^{+  0.03}_{-  0.02}$ &  39.09$^{+  1.06}_{-  1.07}$ &  19.78$^{+  0.03}_{-  0.01}$ &   7.45$^{+  0.15}_{-  0.15}$ &   1.15$^{+  0.02}_{-  0.01}$ &  0.299 \\
12379 &  20.57$^{+  0.08}_{-  0.07}$ &  20.72$^{+  1.05}_{-  0.99}$ &  19.67$^{+  0.17}_{-  0.14}$ &   3.97$^{+  0.40}_{-  0.30}$ &   2.20$^{+  0.20}_{-  0.15}$ &  0.224 \\
12391 &  20.29$^{+  0.07}_{-  0.11}$ &  15.57$^{+  1.46}_{-  1.73}$ &  18.89$^{+  0.21}_{-  0.33}$ &   0.67$^{+  0.09}_{-  0.10}$ &   1.22$^{+  0.16}_{-  0.25}$ &  0.230 \\
12511 &  20.72$^{+  0.04}_{-  0.04}$ &  16.90$^{+  1.54}_{-  1.54}$ &  20.69$^{+  0.17}_{-  0.12}$ &   1.89$^{+  0.22}_{-  0.15}$ &   0.97$^{+  0.17}_{-  0.11}$ &  0.106 \\
12614 &  19.73$^{+  0.01}_{-  0.01}$ &  22.15$^{+  0.34}_{-  0.34}$ &  18.39$^{+  0.02}_{-  0.01}$ &   1.53$^{+  0.02}_{-  0.01}$ &   1.18$^{+  0.02}_{-  0.01}$ &  0.131 \\
12638 &  20.93$^{+  0.03}_{-  0.02}$ &  22.86$^{+  1.16}_{-  1.14}$ &  21.39$^{+  0.15}_{-  0.13}$ &   3.21$^{+  0.32}_{-  0.28}$ &   2.18$^{+  0.14}_{-  0.12}$ &  0.188 \\
12654 &  20.45$^{+  0.04}_{-  0.03}$ &  18.28$^{+  0.75}_{-  0.70}$ &  21.40$^{+  0.21}_{-  0.06}$ &   3.40$^{+  0.53}_{-  0.26}$ &   1.20$^{+  0.18}_{-  0.04}$ &  0.166 \\
12732 & \multicolumn{6}{c}{no fit} \\
12754 &  20.22$^{+  0.07}_{-  0.10}$ &  37.87$^{+  3.55}_{-  4.06}$ &  20.76$^{+  0.02}_{-  0.01}$ &   5.81$^{+  0.44}_{-  0.62}$ &   0.81$^{+  0.03}_{-  0.02}$ &  0.200 \\
12776 &  21.98$^{+  0.06}_{-  0.09}$ &  46.61$^{+  6.70}_{-  6.67}$ &  19.91$^{+  0.08}_{-  0.09}$ &   5.71$^{+  0.24}_{-  0.29}$ &   2.32$^{+  0.10}_{-  0.11}$ &  0.163 \\
12808 &  18.86$^{+  0.05}_{-  0.03}$ &  12.38$^{+  0.25}_{-  0.21}$ &  18.49$^{+  0.46}_{-  0.22}$ &   1.96$^{+  0.51}_{-  0.21}$ &   3.41$^{+  0.72}_{-  0.35}$ &  0.198 \\
12845 &  21.35$^{+  0.01}_{-  0.01}$ &  24.96$^{+  0.61}_{-  0.62}$ &  21.56$^{+  0.03}_{-  0.03}$ &   3.14$^{+  0.07}_{-  0.07}$ &   1.00$^{+  0.03}_{-  0.02}$ &  0.147 \\
\enddata
\tablecomments{See Table 1 for a description of each column.}
\end{deluxetable}

\begin{deluxetable}{crrrrrr}
\tablecolumns{9}
\tablewidth{0pc}
\tablecaption{$B$-band model parameters}
\tablehead{
\colhead{Galaxy} & \colhead{$\mu_{0,d}$} & \colhead{$h$}   &  \colhead{$\mu_{e,b}$}  &
\colhead{$R_{e,b}$} & \colhead{$n$}   & \colhead{extinc} \\
\colhead{UGC \#} & \colhead{mag/$\sq\arcsec$} & \colhead{arcsec}   &  \colhead{mag/$\sq\arcsec$}  &
\colhead{arcsec} & \colhead{}   & \colhead{$B$-mag}
}
\startdata
00089 & \multicolumn{6}{c}{no fit}                                                                                                                                        \\
00093 &  22.35$^{+  0.03}_{-  0.03}$ &  21.63$^{+  1.47}_{-  1.55}$ &  22.16$^{+  0.07}_{-  0.06}$ &   1.34$^{+  0.05}_{-  0.06}$ &   0.50$^{+  0.08}_{-  0.07}$ &  0.246 \\
00242 & \multicolumn{6}{c}{no fit}                                                                                                                                        \\
00334 &  23.37$^{+  0.06}_{-  0.10}$ &  21.19$^{+  3.09}_{-  3.12}$ &  24.82$^{+  0.25}_{-  0.48}$ &   3.20$^{+  1.12}_{-  1.00}$ &   0.99$^{+  0.19}_{-  0.50}$ &  0.240 \\
00438 &  20.58$^{+  0.13}_{-  0.14}$ &  14.33$^{+  0.34}_{-  0.71}$ &  22.97$^{+  0.65}_{-  0.54}$ &   9.72$^{+  7.51}_{-  3.89}$ &   2.71$^{+  0.41}_{-  0.29}$ &  0.153 \\
00463 &  20.76$^{+  0.04}_{-  0.04}$ &  12.99$^{+  0.31}_{-  0.33}$ &  20.51$^{+  0.00}_{-  0.00}$ &   1.57$^{+  0.03}_{-  0.02}$ &   0.37$^{+  0.01}_{-  0.01}$ &  0.392 \\
00490 &  21.31$^{+  0.03}_{-  0.02}$ &  16.38$^{+  0.50}_{-  0.47}$ &  22.03$^{+  0.15}_{-  0.13}$ &   3.41$^{+  0.35}_{-  0.29}$ &   1.56$^{+  0.13}_{-  0.11}$ &  0.215 \\
00508 &  21.69$^{+  0.07}_{-  0.07}$ &  25.58$^{+  1.43}_{-  1.53}$ &  20.29$^{+  0.11}_{-  0.06}$ &   3.82$^{+  0.20}_{-  0.15}$ &   1.38$^{+  0.13}_{-  0.05}$ &  0.298 \\
00628 &  22.78$^{+  0.08}_{-  0.06}$ &  14.28$^{+  1.40}_{-  1.24}$ &  24.19$^{+  0.29}_{-  0.17}$ &   2.85$^{+  0.95}_{-  0.50}$ &   0.67$^{+  0.27}_{-  0.18}$ &  0.190 \\
01305 &  22.25$^{+  0.05}_{-  0.06}$ &  38.06$^{+  2.79}_{-  2.62}$ &  22.37$^{+  0.14}_{-  0.13}$ &   7.98$^{+  0.83}_{-  0.67}$ &   1.99$^{+  0.13}_{-  0.11}$ &  0.307 \\
01455 &  22.37$^{+  0.13}_{-  0.15}$ &  25.18$^{+  3.20}_{-  3.16}$ &  21.29$^{+  0.28}_{-  0.28}$ &   3.79$^{+  0.63}_{-  0.53}$ &   1.70$^{+  0.29}_{-  0.29}$ &  0.445 \\
01551 &  22.55$^{+  0.02}_{-  0.02}$ &  26.93$^{+  1.67}_{-  1.76}$ &  23.95$^{+  0.04}_{-  0.02}$ &   2.40$^{+  0.14}_{-  0.17}$ &   0.57$^{+  0.00}_{-  0.00}$ &  0.391 \\
01559 &  22.57$^{+  0.04}_{-  0.06}$ &  20.60$^{+  1.77}_{-  1.84}$ &  23.08$^{+  0.02}_{-  0.01}$ &   4.71$^{+  0.13}_{-  0.17}$ &   0.58$^{+  0.01}_{-  0.01}$ &  0.266 \\
01577 &  21.83$^{+  0.10}_{-  0.15}$ &  17.59$^{+  2.28}_{-  2.46}$ &  21.15$^{+  0.13}_{-  0.14}$ &   2.77$^{+  0.24}_{-  0.29}$ &   1.20$^{+  0.15}_{-  0.13}$ &  0.247 \\
01719 &  22.65$^{+  0.08}_{-  0.11}$ &  24.13$^{+  2.43}_{-  2.40}$ &  22.25$^{+  0.18}_{-  0.19}$ &   4.47$^{+  0.54}_{-  0.54}$ &   1.94$^{+  0.16}_{-  0.17}$ &  0.351 \\
01792 &  21.71$^{+  0.02}_{-  0.02}$ &  17.57$^{+  0.46}_{-  0.48}$ &  21.38$^{+  0.01}_{-  0.01}$ &   1.91$^{+  0.02}_{-  0.03}$ &   0.53$^{+  0.01}_{-  0.01}$ &  0.385 \\
02064 &  22.45$^{+  0.03}_{-  0.04}$ &  21.67$^{+  1.15}_{-  1.22}$ &  22.00$^{+  0.08}_{-  0.11}$ &   1.76$^{+  0.10}_{-  0.12}$ &   1.32$^{+  0.05}_{-  0.06}$ &  0.570 \\
02081 &  22.35$^{+  0.07}_{-  0.08}$ &  19.44$^{+  2.43}_{-  2.14}$ &  23.65$^{+  0.05}_{-  0.05}$ &   2.39$^{+  0.50}_{-  0.42}$ &   0.67$^{+  0.08}_{-  0.08}$ &  0.112 \\
02124 &  22.19$^{+  0.06}_{-  0.06}$ &  21.35$^{+  1.48}_{-  1.54}$ &  20.65$^{+  0.01}_{-  0.01}$ &   4.58$^{+  0.04}_{-  0.04}$ &   0.71$^{+  0.02}_{-  0.02}$ &  0.139 \\
02125 &  23.08$^{+  0.02}_{-  0.04}$ &  26.33$^{+  1.73}_{-  1.76}$ &  21.85$^{+  0.04}_{-  0.05}$ &   2.84$^{+  0.08}_{-  0.07}$ &   1.49$^{+  0.08}_{-  0.09}$ &  0.610 \\
02197 &  22.58$^{+  0.02}_{-  0.03}$ &  18.08$^{+  0.85}_{-  0.84}$ &  23.41$^{+  0.10}_{-  0.09}$ &   2.77$^{+  0.23}_{-  0.19}$ &   0.73$^{+  0.10}_{-  0.09}$ &  0.927 \\
02368 &  22.04$^{+  0.02}_{-  0.01}$ &  18.84$^{+  0.45}_{-  0.46}$ &  19.03$^{+  0.10}_{-  0.10}$ &   0.82$^{+  0.03}_{-  0.04}$ &   2.49$^{+  0.23}_{-  0.22}$ &  0.721 \\
02595 & \multicolumn{6}{c}{no data}                                                                                                                                       \\
03066 &  22.03$^{+  0.06}_{-  0.07}$ &  14.09$^{+  0.80}_{-  0.90}$ &  22.84$^{+  0.07}_{-  0.16}$ &   1.22$^{+  0.16}_{-  0.19}$ &   0.52$^{+  0.05}_{-  0.16}$ &  1.332 \\
03080 &  22.19$^{+  0.03}_{-  0.03}$ &  19.50$^{+  1.52}_{-  1.63}$ &  22.42$^{+  0.04}_{-  0.06}$ &   1.39$^{+  0.06}_{-  0.08}$ &   0.32$^{+  0.04}_{-  0.08}$ &  0.372 \\
03140 &  21.05$^{+  0.16}_{-  0.14}$ &  13.71$^{+  0.49}_{-  0.77}$ &  22.16$^{+  1.35}_{-  0.83}$ &   4.83$^{+  6.53}_{-  1.94}$ &   2.87$^{+  1.28}_{-  0.82}$ &  0.349 \\
04126 &  21.69$^{+  0.05}_{-  0.08}$ &  20.81$^{+  1.30}_{-  1.47}$ &  20.86$^{+  0.06}_{-  0.09}$ &   2.27$^{+  0.12}_{-  0.15}$ &   1.16$^{+  0.04}_{-  0.05}$ &  0.224 \\
04256 &  21.32$^{+  0.01}_{-  0.00}$ &  20.49$^{+  0.40}_{-  0.39}$ &  20.10$^{+  0.04}_{-  0.03}$ &   1.07$^{+  0.01}_{-  0.01}$ &   1.47$^{+  0.09}_{-  0.07}$ &  0.231 \\
04308 &  21.47$^{+  0.02}_{-  0.02}$ &  22.21$^{+  0.84}_{-  0.88}$ &  20.55$^{+  0.03}_{-  0.04}$ &   1.48$^{+  0.03}_{-  0.03}$ &   0.64$^{+  0.03}_{-  0.03}$ &  0.168 \\
04368 &  21.54$^{+  0.06}_{-  0.08}$ &  17.66$^{+  1.16}_{-  1.34}$ &  22.17$^{+  0.01}_{-  0.00}$ &   2.29$^{+  0.14}_{-  0.20}$ &   0.42$^{+  0.03}_{-  0.04}$ &  0.162 \\
04375 &  21.31$^{+  0.08}_{-  0.12}$ &  19.86$^{+  1.62}_{-  1.89}$ &  22.19$^{+  0.17}_{-  0.21}$ &   2.62$^{+  0.57}_{-  0.60}$ &   0.93$^{+  0.13}_{-  0.16}$ &  0.194 \\
04422 &  22.14$^{+  0.03}_{-  0.07}$ &  32.40$^{+  3.43}_{-  3.87}$ &  20.48$^{+  0.02}_{-  0.02}$ &   3.50$^{+  0.04}_{-  0.05}$ &   0.46$^{+  0.02}_{-  0.02}$ &  0.172 \\
04458 &  22.91$^{+  0.21}_{-  0.27}$ &  29.09$^{+  3.81}_{-  3.67}$ &  21.13$^{+  0.16}_{-  0.21}$ &   7.51$^{+  0.72}_{-  0.88}$ &   2.70$^{+  0.17}_{-  0.22}$ &  0.150 \\
05103 &  20.51$^{+  0.06}_{-  0.09}$ &  17.36$^{+  0.94}_{-  1.14}$ &  19.85$^{+  0.05}_{-  0.06}$ &   1.77$^{+  0.08}_{-  0.11}$ &   0.65$^{+  0.05}_{-  0.06}$ &  0.114 \\
05303 &  21.25$^{+  0.02}_{-  0.03}$ &  32.83$^{+  0.86}_{-  0.91}$ &  21.32$^{+  0.01}_{-  0.01}$ &   3.71$^{+  0.08}_{-  0.09}$ &   0.73$^{+  0.00}_{-  0.01}$ &  0.148 \\
05510 &  20.67$^{+  0.04}_{-  0.04}$ &  20.30$^{+  0.73}_{-  0.80}$ &  19.53$^{+  0.07}_{-  0.08}$ &   1.61$^{+  0.06}_{-  0.07}$ &   1.18$^{+  0.08}_{-  0.08}$ &  0.100 \\
05554 &  21.06$^{+  0.05}_{-  0.05}$ &  19.12$^{+  0.77}_{-  0.82}$ &  19.56$^{+  0.06}_{-  0.08}$ &   1.95$^{+  0.06}_{-  0.07}$ &   1.08$^{+  0.07}_{-  0.08}$ &  0.116 \\
05633 &  23.00$^{+  0.34}_{-  0.19}$ &  26.15$^{+  9.07}_{-  5.54}$ &  23.78$^{+  1.04}_{-  0.30}$ &   5.05$^{+  7.92}_{-  1.43}$ &   0.84$^{+  0.78}_{-  0.29}$ &  0.186 \\
05842 &  21.95$^{+  0.00}_{-  0.01}$ &  55.99$^{+  0.53}_{-  0.53}$ &  21.75$^{+  0.00}_{-  0.00}$ &   2.75$^{+  0.01}_{-  0.00}$ &   0.48$^{+  0.00}_{-  0.00}$ &  0.121 \\
06028 & \multicolumn{6}{c}{no fit}                                                                                                                                        \\
06077 &  21.06$^{+  0.05}_{-  0.08}$ &  18.77$^{+  1.15}_{-  1.33}$ &  20.42$^{+  0.05}_{-  0.08}$ &   1.50$^{+  0.08}_{-  0.10}$ &   0.80$^{+  0.04}_{-  0.05}$ &  0.092 \\
06123 &     NP                       &  26.10$^{+  2.29}_{-  2.84}$ &     NP                       &   2.03$^{+  0.33}_{-  0.54}$ &   1.12$^{+  0.08}_{-  0.11}$ &  0.105 \\
06277 &  20.76$^{+  0.01}_{-  0.03}$ &  24.98$^{+  0.61}_{-  0.62}$ &  19.61$^{+  0.07}_{-  0.07}$ &   1.65$^{+  0.06}_{-  0.06}$ &   1.59$^{+  0.09}_{-  0.08}$ &  0.095 \\
06445 &  20.65$^{+  0.32}_{-  0.12}$ &  16.73$^{+  1.83}_{-  1.08}$ &  20.26$^{+  1.06}_{-  0.16}$ &   3.91$^{+  3.39}_{-  0.38}$ &   1.04$^{+  0.96}_{-  0.15}$ &  0.113 \\
06453 &  20.39$^{+  0.01}_{-  0.02}$ &  16.84$^{+  0.30}_{-  0.31}$ &  20.98$^{+  0.03}_{-  0.03}$ &   2.55$^{+  0.07}_{-  0.06}$ &   0.61$^{+  0.03}_{-  0.03}$ &  0.111 \\
06460 &  20.66$^{+  0.03}_{-  0.02}$ &  26.51$^{+  0.72}_{-  0.77}$ &  19.51$^{+  0.03}_{-  0.03}$ &   2.05$^{+  0.04}_{-  0.04}$ &   0.74$^{+  0.03}_{-  0.02}$ &  0.105 \\
06536 &     NP                       &  24.54$^{+  6.11}_{-  6.03}$ &     NP                       &   5.67$^{+  0.79}_{-  1.11}$ &   1.89$^{+  0.19}_{-  0.33}$ &  0.101 \\
06693 &  21.88$^{+  0.00}_{-  0.00}$ &  23.32$^{+  0.64}_{-  0.65}$ &  22.05$^{+  0.01}_{-  0.01}$ &   1.27$^{+  0.02}_{-  0.02}$ &   0.63$^{+  0.00}_{-  0.01}$ &  0.105 \\
06746 &  21.46$^{+  0.16}_{-  0.10}$ &  18.09$^{+  2.63}_{-  1.63}$ &  20.51$^{+  0.18}_{-  0.07}$ &   2.95$^{+  0.35}_{-  0.16}$ &   1.07$^{+  0.22}_{-  0.08}$ &  0.105 \\
06754 &  22.96$^{+  0.01}_{-  0.14}$ &  39.92$^{+  2.57}_{-  6.16}$ &  21.80$^{+  0.02}_{-  0.28}$ &   4.58$^{+  0.06}_{-  0.20}$ &   1.19$^{+  0.04}_{-  0.61}$ &  0.116 \\
07169 &  19.94$^{+  0.03}_{-  0.02}$ &  12.71$^{+  0.32}_{-  0.32}$ &  20.12$^{+  0.07}_{-  0.05}$ &   1.69$^{+  0.08}_{-  0.06}$ &   0.90$^{+  0.07}_{-  0.05}$ &  0.145 \\
07315 &  20.08$^{+  0.05}_{-  0.18}$ &  15.16$^{+  0.66}_{-  1.53}$ &  20.76$^{+  0.19}_{-  0.34}$ &   2.10$^{+  0.34}_{-  0.58}$ &   0.81$^{+  0.16}_{-  0.30}$ &  0.130 \\
07450 &  21.14$^{+  0.05}_{-  0.06}$ &  62.02$^{+  3.30}_{-  3.64}$ &  19.48$^{+  0.01}_{-  0.01}$ &   6.88$^{+  0.04}_{-  0.06}$ &   0.37$^{+  0.02}_{-  0.02}$ &  0.113 \\
07523 &  21.22$^{+  0.07}_{-  0.07}$ &  32.05$^{+  1.72}_{-  1.80}$ &  19.96$^{+  0.05}_{-  0.05}$ &   5.50$^{+  0.18}_{-  0.21}$ &   1.27$^{+  0.04}_{-  0.05}$ &  0.131 \\
07594 &  21.34$^{+  0.21}_{-  0.21}$ &  57.14$^{+  4.43}_{-  5.02}$ &  22.40$^{+  0.81}_{-  0.73}$ &  30.23$^{+ 20.86}_{- 11.45}$ &   3.94$^{+  0.70}_{-  0.63}$ &  0.120 \\
07876 &     NP                       &  19.82$^{+  1.27}_{-  1.58}$ &      NP                      &   1.66$^{+  0.24}_{-  0.34}$ &   0.43$^{+  0.04}_{-  0.06}$ &  0.117 \\
07901 &  19.77$^{+  0.01}_{-  0.00}$ &  20.47$^{+  0.22}_{-  0.21}$ &  20.04$^{+  0.02}_{-  0.02}$ &   2.69$^{+  0.03}_{-  0.04}$ &   1.24$^{+  0.02}_{-  0.02}$ &  0.115 \\
08279 & \multicolumn{6}{c}{no fit}                                                                                                                                        \\
08289 &  21.81$^{+  0.01}_{-  0.01}$ &  34.59$^{+  1.14}_{-  1.26}$ &  19.16$^{+  0.00}_{-  0.00}$ &   2.89$^{+  0.00}_{-  0.01}$ &   0.55$^{+  0.01}_{-  0.00}$ &  0.103 \\
08865 &  21.79$^{+  0.11}_{-  0.17}$ &  29.37$^{+  2.90}_{-  3.27}$ &  20.71$^{+  0.09}_{-  0.11}$ &   5.26$^{+  0.35}_{-  0.44}$ &   1.30$^{+  0.08}_{-  0.10}$ &  0.089 \\
09024 &  24.37$^{+  0.00}_{-  0.00}$ &  32.84$^{+  0.01}_{-  0.00}$ &  22.13$^{+  0.00}_{-  0.00}$ &   2.75$^{+  0.00}_{-  0.00}$ &   1.34$^{+  0.00}_{-  0.00}$ &  0.117 \\
09061 &  22.66$^{+  0.01}_{-  0.02}$ &  60.97$^{+  4.52}_{-  4.72}$ &  20.65$^{+  0.02}_{-  0.02}$ &   3.67$^{+  0.03}_{-  0.04}$ &   0.98$^{+  0.03}_{-  0.02}$ &  0.103 \\
09481 &  21.33$^{+  0.04}_{-  0.05}$ &  18.63$^{+  0.90}_{-  0.99}$ &  21.41$^{+  0.03}_{-  0.04}$ &   1.92$^{+  0.09}_{-  0.11}$ &   0.66$^{+  0.03}_{-  0.03}$ &  0.074 \\
09915 &      NP                      &  18.43$^{+  1.39}_{-  1.57}$ &     NP                       &   3.49$^{+  0.17}_{-  0.22}$ &   0.63$^{+  0.02}_{-  0.03}$ &  0.186 \\
09926 &  20.12$^{+  0.59}_{-  0.11}$ &  17.91$^{+  4.43}_{-  1.35}$ &  20.16$^{+  1.88}_{-  0.24}$ &   3.43$^{+  9.80}_{-  0.47}$ &   1.12$^{+  1.48}_{-  0.25}$ &  0.235 \\
09943 &  20.30$^{+  0.23}_{-  0.31}$ &  19.25$^{+  2.53}_{-  3.00}$ &  21.05$^{+  0.21}_{-  0.24}$ &   5.63$^{+  1.80}_{-  1.88}$ &   0.97$^{+  0.17}_{-  0.25}$ &  0.179 \\
10083 &  20.87$^{+  0.08}_{-  0.07}$ &  17.64$^{+  1.01}_{-  0.97}$ &  23.75$^{+  0.08}_{-  0.19}$ &   7.61$^{+  2.27}_{-  2.23}$ &   2.11$^{+  0.00}_{-  0.02}$ &  0.157 \\
10437 &  \multicolumn{6}{c}{no fit}                                                                                                                                       \\
10445 &  22.08$^{+  0.03}_{-  0.06}$ &  26.13$^{+  3.62}_{-  3.84}$ &  22.77$^{+  0.03}_{-  0.01}$ &   4.01$^{+  0.18}_{-  0.27}$ &   0.51$^{+  0.02}_{-  0.03}$ &  0.139 \\
10584 &  21.89$^{+  0.02}_{-  0.03}$ &  23.94$^{+  1.64}_{-  1.74}$ &  20.91$^{+  0.01}_{-  0.02}$ &   1.86$^{+  0.02}_{-  0.03}$ &   0.38$^{+  0.02}_{-  0.02}$ &  0.055 \\
11628 &  22.81$^{+  0.08}_{-  0.11}$ &  50.63$^{+  7.58}_{-  7.57}$ &  21.45$^{+  0.10}_{-  0.13}$ &   7.75$^{+  0.49}_{-  0.60}$ &   2.18$^{+  0.09}_{-  0.12}$ &  0.423 \\
11708 &  21.51$^{+  0.05}_{-  0.06}$ &  17.73$^{+  1.27}_{-  1.36}$ &  21.68$^{+  0.05}_{-  0.05}$ &   2.05$^{+  0.13}_{-  0.15}$ &   0.82$^{+  0.04}_{-  0.05}$ &  0.291 \\
11872 &  20.81$^{+  0.49}_{-  0.42}$ &  20.37$^{+  2.87}_{-  2.34}$ &  20.96$^{+  0.12}_{-  0.14}$ &  12.26$^{+  2.30}_{-  2.39}$ &   1.49$^{+  0.09}_{-  0.11}$ &  0.310 \\
12151 &  23.10$^{+  0.00}_{-  0.13}$ &  23.47$^{+  2.55}_{-  6.63}$ &  23.83$^{+  0.28}_{-  0.00}$ &   3.84$^{+  0.14}_{-  0.24}$ &   0.15$^{+  0.00}_{-  0.00}$ &  0.300 \\
12343 &  21.45$^{+  0.03}_{-  0.04}$ &  39.29$^{+  1.55}_{-  1.71}$ &  21.46$^{+  0.02}_{-  0.04}$ &   6.52$^{+  0.20}_{-  0.21}$ &   0.95$^{+  0.03}_{-  0.05}$ &  0.482 \\
12379 &  22.15$^{+  0.20}_{-  0.24}$ &  20.25$^{+  3.27}_{-  3.15}$ &  21.35$^{+  0.46}_{-  0.37}$ &   3.61$^{+  1.11}_{-  0.73}$ &   1.89$^{+  0.47}_{-  0.36}$ &  0.362 \\
12391 &  21.68$^{+  0.06}_{-  0.11}$ &  16.32$^{+  1.54}_{-  1.92}$ &  19.73$^{+  0.18}_{-  0.33}$ &   0.56$^{+  0.02}_{-  0.02}$ &   0.56$^{+  0.26}_{-  0.41}$ &  0.371 \\
12511 &  22.43$^{+  0.01}_{-  0.02}$ &  23.59$^{+  0.87}_{-  0.90}$ &  22.22$^{+  0.05}_{-  0.04}$ &   1.61$^{+  0.05}_{-  0.05}$ &   0.75$^{+  0.04}_{-  0.03}$ &  0.171 \\
12614 &  20.98$^{+  0.01}_{-  0.02}$ &  21.19$^{+  0.46}_{-  0.48}$ &  19.23$^{+  0.02}_{-  0.02}$ &   1.09$^{+  0.01}_{-  0.01}$ &   0.73$^{+  0.02}_{-  0.03}$ &  0.211 \\
12638 &  22.48$^{+  0.00}_{-  0.01}$ &  28.33$^{+  1.67}_{-  1.69}$ &  22.94$^{+  0.09}_{-  0.10}$ &   3.13$^{+  0.20}_{-  0.20}$ &   1.91$^{+  0.08}_{-  0.08}$ &  0.303 \\
12654 &  21.89$^{+  0.01}_{-  0.02}$ &  21.80$^{+  0.81}_{-  0.82}$ &  22.94$^{+  0.03}_{-  0.02}$ &   3.13$^{+  0.12}_{-  0.12}$ &   1.05$^{+  0.02}_{-  0.02}$ &  0.268 \\
12732 & \multicolumn{6}{c}{no fit}                                                                                                                                        \\
12754 &  21.48$^{+  0.04}_{-  0.05}$ &  42.37$^{+  2.97}_{-  3.20}$ &  21.76$^{+  0.02}_{-  0.02}$ &   5.72$^{+  0.26}_{-  0.31}$ &   0.87$^{+  0.02}_{-  0.02}$ &  0.324 \\
12776 &  23.54$^{+  0.00}_{-  0.01}$ &  74.88$^{+  6.50}_{-  6.37}$ &  21.69$^{+  0.04}_{-  0.04}$ &   6.19$^{+  0.13}_{-  0.14}$ &   2.27$^{+  0.05}_{-  0.05}$ &  0.263 \\
12808 &  20.17$^{+  0.02}_{-  0.02}$ &  12.57$^{+  0.25}_{-  0.26}$ &  18.20$^{+  0.03}_{-  0.04}$ &   0.92$^{+  0.01}_{-  0.02}$ &   1.30$^{+  0.04}_{-  0.04}$ &  0.320 \\
12845 &  22.77$^{+  0.02}_{-  0.03}$ &  25.36$^{+  1.40}_{-  1.45}$ &  23.18$^{+  0.02}_{-  0.02}$ &   2.92$^{+  0.10}_{-  0.11}$ &   0.68$^{+  0.03}_{-  0.02}$ &  0.237 \\
\enddata
\tablecomments{See Table 1 for a description of each column.}
\end{deluxetable}

\end{document}